\documentclass{jfm}
\usepackage{graphicx}
\usepackage{natbib}

\ifCUPmtlplainloaded \else
  \checkfont{eurm10}
  \iffontfound
    \IfFileExists{upmath.sty}
      {\typeout{^^JFound AMS Euler Roman fonts on the system,
                   using the 'upmath' package.^^J}%
       \usepackage{upmath}}
      {\typeout{^^JFound AMS Euler Roman fonts on the system, but you
                   dont seem to have the}%
       \typeout{'upmath' package installed. JFM.cls can take advantage
                 of these fonts,^^Jif you use 'upmath' package.^^J}%
       \providecommand\upi{\pi}%
      }
  \else
    \providecommand\upi{\pi}%
  \fi
\fi

\ifCUPmtlplainloaded \else
  \checkfont{msam10}
  \iffontfound
    \IfFileExists{amssymb.sty}
      {\typeout{^^JFound AMS Symbol fonts on the system, using the
                'amssymb' package.^^J}%
       \usepackage{amssymb}%

      }{}
  \fi
\fi

\ifCUPmtlplainloaded \else
  \IfFileExists{amsbsy.sty}
    {\typeout{^^JFound the 'amsbsy' package on the system, using it.^^J}%
     \usepackage{amsbsy}}
    {\providecommand\boldsymbol[1]{\mbox{\boldmath $##1$}}}
\fi

\newcommand\Rey{\mbox{\textit{Re}}}  % Reynolds number
\newsavebox{\astrutbox}
\sbox{\astrutbox}{\rule[-5pt]{0pt}{20pt}}

\newcommand\p{\ensuremath{\partial}}

\newcommand\eg{e.g.\ }

\title[LES               of large-scale structures in long channel flow]
  {Large-eddy simulation of large-scale structures in long channel flow}

\author[D. Chung and B. J. McKeon]%
{D.\ns C\ls H\ls U\ls N\ls G \and B.\ns J.\ns M\ls c\ls K\ls E\ls O\ls N\thanks{Email address for correspondence: mckeon@caltech.edu}}

\affiliation{Graduate Aerospace Laboratories, California Institute of
    Technology, Pasadena, CA 91125, USA\break
    dchung@caltech.edu}

\pubyear{1996}
\volume{538}
\pagerange{119--126}
\date{?? and in revised form ??}

\usepackage{amsmath}

\newcommand\bm\boldsymbol
\def\drawline#1#2{\raise 2.5pt\vbox{\hrule width #1pt height #2pt}}
\def\spacce#1{\hskip #1pt}
\def\solid{\drawline{24}{.5}}
\def\bdash{\hbox{\drawline{4}{.5}\spacce{2}}}
\def\dashed{\bdash\bdash\bdash\bdash}
\usepackage{color}

\begin{document}

\maketitle

\begin{abstract}
We investigate statistics of large-scale structures from large-eddy
simulation (LES) of turbulent channel flow at friction Reynolds numbers
$\Rey_\tau = 2\,{\rm k}$ and $200\,{\rm k}$.
To properly capture the behaviour of large-scale structures,
the channel length is chosen to be 96 times the channel half-height.
In agreement with experiments, these large-scale structures are found
to give rise to an apparent amplitude modulation of the underlying small-scale fluctuations. This effect is explained in terms of the phase relationship between the large- and small-scale activity.  The shape of the dominant large-scale structure is investigated by conditional averages based on the large-scale velocity, determined using a filter width equal to the channel half-height. The conditioned field demonstrates coherence on a scale of several times the filter width, and the small-scale--large-scale relative phase difference increases away from the wall, passing through $\upi/2$ in the overlap region of the mean velocity before approaching $\upi$ further from the wall. We also found that, near the wall, the convection velocity of the large-scales departs slightly, but unequivocally, from the mean velocity.
\end{abstract}

\section{Introduction}

Recent studies \cite*[]{Kim99, Morrison04, Guala2006, Monty2007, Hutchins2007a, Hutchins2007b,
Mathis2009} have confirmed earlier observations \cite*[]{Favre67, Kovasznay70} of very long large-scale structures in
the wall region of boundary layers, channels and pipes.
These structures are marked by streamwise-elongated, alternating
low- and high-momentum, meandering streaks, with length
$10 \delta$ \cite[]{Kim99,Morrison04}, $8$--$16\delta$ \cite[]{Guala2006}, $25\delta$ \cite[]{Monty2007}, $6\delta$ \cite[]{Hutchins2007a},
$20\delta$ \cite[]{Hutchins2007b} and width
$0.3$--$0.5\delta$ \cite[]{Mathis2009}, where $\delta$ is the boundary layer
thickness, channel half-height or pipe diameter. See \cite{Monty09} for a description of the differences between the characteristics of these large structures in the different canonical flows.
The bursting period of these structures,
$6\delta/U$, where $U(z)$ is the mean velocity,
was already noted some decades ago, along with the long tails of the streamwise velocity auto-correlations, see review by \cite{Cantwell1981}.
The dynamical significance of these large-scale structures can be
seen in a scale decomposition of relative energy content, as measured by
the premultiplied one-dimensional
longitudinal spectrum $\kappa_x E_{uu}$ plotted against the log
streamwise wavelength,
$\log\lambda_x$, where $\lambda_x = 2\upi/\kappa_x$ (equal area under
the curve implies equal energy contribution).
For a boundary layer at friction Reynolds number
$\Rey_\tau = 7.3\,{\rm k}$ \cite[]{Hutchins2007a},
the signature of these structures are related to the outer peak
in $\kappa_x E_{uu}$ found at $(z/\delta,\lambda_x/\delta)=(0.06,6)$.
It has been proposed that the wall-normal location of this peak is located
at the middle of the log layer, $z^+\propto \Rey_\tau^{1/2}$ or $\Rey_\tau^{3/4}$ \cite[]{Mathis2009}
(the choice of scaling depends on whether the lower limit of the log law is Reynolds number dependent),
where $z$ is the height from the wall and the superscript $+$ indicates
scaling in wall units: the friction velocity $u_\tau$ and kinematic
viscosity $\nu$. However the scaling remains somewhat ambiguous.

The large-scale structures were found \cite[]{Bandyopadhyay1984,Mathis2009} to
modulate the amplitudes of superimposed small-scale fluctuations.
To test this idea, these authors first split the streamwise velocity
into large- and small-scale components via a
filter at $\lambda_x/\delta = U(z)/(f \delta)$,
and then used either a filtered and rectified small-scale signal or the Hilbert-transform to determine the envelope for the small-scale fluctuations,
finally forming the correlation coefficient between the large-scale
fluctuations and the low-pass filtered envelope of the small-scale fluctuations.
They found that, near the wall, large-scale high-speed
regions carry intense superimposed small-scale fluctuations, but
this correlation is reversed above a height that
decreases in outer units with $\Rey_\tau$.
We shall attempt to reproduce these features presently.

The footprint of structures centred far from the wall provides an obvious challenge in terms of determining appropriate convection velocities across the range of turbulent scales, with particular importance for obtaining the correct wavenumber spectra from temporal frequency spectra obtained by, for example, hot-wire anemometry. It has been known for some time that convection velocities deviate from the local mean in the near-wall region, \cite[\eg][]{Krogstad98_Taylor}. The common practice is to use Taylor's frozen-turbulence hypothesis to map from the frequency to the wavenumber domain, that is
to use the assumption that all structures at a given wall distance $z$ convect
at the same scale-independent mean velocity $U(z)$.
It was shown from a particle image velocimetry (PIV) experiment
\cite[]{Dennis2008} that this is indeed good approximation
at $z/\delta = 0.16$ for a $\Rey_\theta = 4.7\,{\rm k}$ boundary layer,
at least for scales smaller than their field of view, $3.2\delta$ in space
and $6.3\delta/U$ in time. However note that this wall-normal distance is sufficiently far from the wall that it is beyond the large-scale energy peak, such that any convection velocity questions are likely insignificant because of the low shear in the outer region.
With a field of view larger than $20\delta \times 20 \delta/U$ and
height down to $z/\delta = 0.049$, we
revisit the question of
whether the footprint of the large-scale structures, having centres further
from the wall, still convect at the local mean velocity near the wall.

To properly assess the dynamics of these long structures, reported
to reach up to $25\delta$ \cite[]{Monty2007},
we use large-eddy simulation (LES) coupled with a wall model
\cite[]{Chung2009}.
This investigation is ideally suited to the present wall modelled LES
since its cost depends only on the number of `large eddies', which,
for a channel,
is Reynolds number independent.
In contrast, the fully resolved direct numerical simulation (DNS)
is prohibitively expensive.
For reference, the most ambitious DNS of a channel flow to date
is the $\Rey_\tau = 2\,{\rm k}$
simulation \cite[]{Hoyas2006} in an $L_x/\delta = 8\upi \approx 25$ channel,
where $L_x$ is the streamwise length;
a DNS investigation at higher $\Rey_\tau$ and larger $L_x$
of these large-scale structures is not yet possible.
Of course, the use of LES comes at the cost of subgrid-scale (SGS) modelling,
wall modelling and numerical errors,
but LES is
much faster (hours for simulation, minutes for post-processing)
than DNS and experiments; we
hope that a controlled application of the present LES combined with
experience in the subject may shed some light on the physics of these
large-scale structures.

Details of the simulations are given in \S\,\ref{sec:simulation} and
discussion of observations are found in \S\,\ref{sec:discussion} before
we conclude in \S\,\ref{sec:conclusions}.

\section{Simulation details}
\label{sec:simulation}

As full details of the LES, including the numerical method and
SGS model, are given by \cite{Chung2009}, we only highlight the
important points here.
We solve the filtered Navier--Stokes equations for the LES velocity
field $\overline{{\bm u}}$ using the stretched-spiral vortex SGS model
\cite[]{Misra1997, Voelkl2000}.
To circumvent the inhibitive cost of resolving the near-wall region
\cite[]{Chapman1979}, $z < h_0$, we use a wall model \cite[]{Chung2009} that
supplies off-wall slip-velocity boundary conditions at $h_0$ to the
interior LES, operating in $h_0 < z < 2 \delta - h_0$,
where $z = 0, 2\delta$ locates the walls.
Presently, we fix $h_0 = 0.18\,\Delta_z$, and the slip velocity is calculated
using a wall model comprising 1) an evolution
equation for the wall shear stress derived from assuming local inner scaling for the
streamwise momentum equation and 2) an extended form of the
stretched-vortex SGS model that provides a local log relation
along with a dynamic estimate for the local K\'arm\'an constant.

\begin{table}
  \begin{center}
\def~{\hphantom{0}}
  \begin{tabular}{cr@{\,}ccccccccccc}
Run & \multicolumn{2}{c}{$\Rey_\tau$} & $L_x/\delta$ & $L_y/\delta$ &
  $h_0^+$ & $\Delta_x/\delta$ & $\Delta_t u_\tau/\delta$ &
  $N_x$ & $N_y$ & $N_z$ & $N_t$ &
  $T U_c/L_x$\\[3pt]
G1\hphantom{b} &    2 & k & 95 & 7.9 & \hphantom{7}15 & 0.17\hphantom{0} & 0.006 &
  \hphantom{1}576 & 48 & 48 & 72000 & 110 \\
H3\hphantom{b} &  200 & k & 96 & 8.0 &            750 & 0.083 & 0.002 &
             1152 & 96 & 96 & 15800 & \hphantom{1}12 \\
G1b            &    2 & k & 95 & 7.9 & \hphantom{7}15 & 0.17\hphantom{0} & 0.006 &
  \hphantom{1}576 & 48 & 48 & 22000 & \hphantom{1}34 \\
\end{tabular}
  \caption{LES parameters for long channel flows. Channel-transit times based on data-recording period
       $T$ and centreline velocity $U_c$.}
  \label{tab:runs2}
   \end{center}
\end{table}
The parameters for the three LES runs are given in
table \ref{tab:runs2}.
The grid is uniform, $\Delta_x = \Delta_y = 4\Delta_z$, throughout the
simulation domain.
To capture the physics of long large-scale structures, we use long a
channel, $L_x/\delta\approx96$, and the statistics are taken over
$T$ channel-transit times.

\begin{figure}
\centerline{\includegraphics{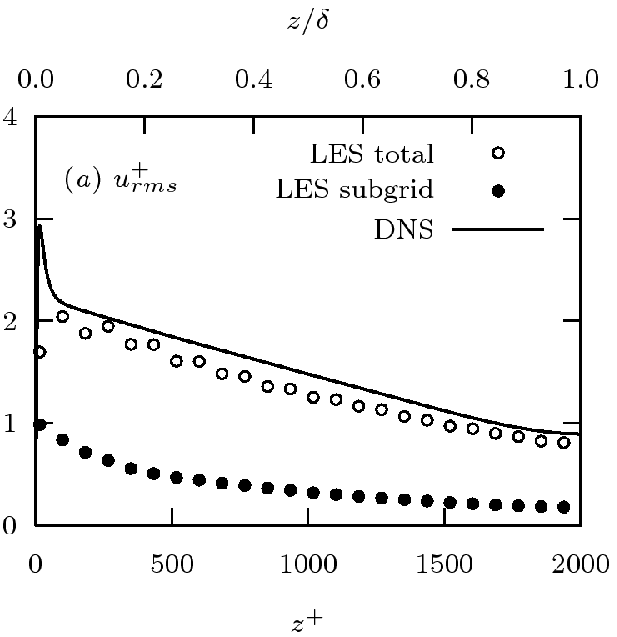}
            \includegraphics{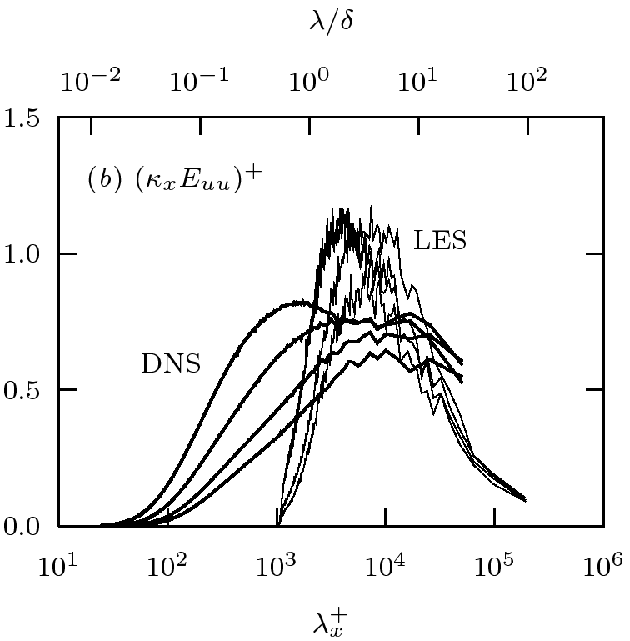}}
\caption{Comparison between DNS data of \cite{Hoyas2006} and present
         $\Rey = 2\,{\rm k}$ channel flow LES, run G1b (table \ref{tab:runs2}).
         Spectra at $z/\delta = 0.049, 0.090, 0.17, 0.26
         \Leftrightarrow z^+ = 98, 180, 350, 510$ (ordered in decreasing energy).}
\label{fig:compareDns2k}
\end{figure}
It was shown \cite[]{Chung2009} that the LES-predicted statistics from
the $\Rey = 2\,{\rm k}$ case, including
means, turbulent intensities and spectra, compare reasonably well with the DNS of
\cite{Hoyas2006}.
We show the root-mean-square (r.m.s.) of the streamwise velocity
fluctuations and the LES-resolved spectra in figure \ref{fig:compareDns2k}.
The points in \ref{fig:compareDns2k}(\textit{a}) correspond to actual
discretization points.
Even though the total (subgrid plus resolved) r.m.s.\ is within $90\%$ of
the DNS result, its spectra plotted in energy-content form overpredicts
the DNS spectra by about $20\%$. (Note, however, that the comparison between experimental channel flow data and DNS has revealed a similar trend, with the experimental premultiplied spectra at large wavelengths in the overlap layer being up to order $10\%$ larger than the equivalent DNS values \cite[]{Monty2009}.)
As such, the results presented here should be viewed as approximate, despite capturing energy at significantly larger wavelengths.
On the other hand, the physics reported here can be seen as robust features
of wall turbulence if they are also observed elsewhere.
We note that the peak value, $(\kappa_x E_{uu})^+ \approx 1$ at $z/\delta = 0.090$,
$\lambda/\delta \approx 6$ (figure \ref{fig:compareDns2k})
is within the range of the peak values from boundary layer
spectra for $\Rey_\tau = 1.0$--$7.3\,{\rm k}$ (see figure 9 of \cite{Hutchins2007b}).
We interpret the LES results as a model of the real flow, and emphasize that $\Rey_\tau = 200 {\rm k}$ is far out of the reach of current DNS approaches.

In order to compute correlations based on temporal
averaging, a numerical rake in runs G1 and H3, fixed in streamwise--spanwise
location, is set up to record the
LES velocity $\overline{u}$ and its modelled subgrid fluctuations
$T_{xx}$ ($\equiv \overline{u u} - \overline{u}\,\overline{u}$)
at the wall-normal locations $z = n_z \Delta_z$
($n_z = 0, 1, \dots, N_z$) and time steps
$t = n_t \Delta_t$ ($n_t = 0, 1, \dots, N_t-1$).
Analogous correlations based on spatial averages are also computed
from these runs from a snapshot in time.

The only difference between runs G1 and G1b is the recorded data.
For the latter, the three-dimensional data set,
$\overline{u}(n_x \Delta_x, y, n_z \Delta_z, n_t \Delta_t)$,
is recorded at fixed $y$,
for $n_x = 0, 1, \dots, 575$, $n_z = 0, 1,\dots,48$
and $n_t = 0, 1,\dots,21999$, where
$\Delta_x/\delta = 0.17$, $\Delta_z/\delta =  0.041$
and $\Delta_t u_\tau/\delta = 0.006$.

Presently, $x$, $y$ and $z$ respectively denote the streamwise,
spanwise and wall-normal directions;
the velocity components, $u$, $v$ and $w$, are defined accordingly.

\section{Discussion of observations}
\label{sec:discussion}

We present observations of the LES velocity fields, with emphasis on the large scales.

\subsection{Convection velocities from spatio-temporal spectra}

We begin by using the somewhat unique combination of spatial and temporal data available in this study to investigate the validity of Taylor's hypothesis. Given the autocorrelation of the streamwise velocity fluctuations,
\begin{equation}
\label{eq:autocorrelation}
R(\rho, \tau) = \langle u'(x,t)u'(x+\rho,t+\tau)\rangle,
\end{equation}
where $\rho$ is the streamwise separation;
$\tau$ is the time delay; and
$u \equiv U + u'$ such that $\langle u' \rangle = 0$,
we define the spatio-temporal
spectrum $\Psi(\kappa,\omega)$ as the Fourier transform of
$R(\rho, \tau)$.
That is, together they form the Fourier transform pair, given by
\begin{eqnarray*}
\Psi(\kappa,\omega) &=& \dfrac{1}{(2\upi)^2}
\int_{-\infty}^\infty \int_{-\infty}^\infty
   R(\rho,\tau) \,{\rm e}^{-{\rm i}(\kappa\rho-\omega\tau)}
  \,{\rm d}\rho\,{\rm d}\tau, \\
R(\rho,\tau) &=&
\int_{-\infty}^\infty \int_{-\infty}^\infty
   \Psi(\kappa,\omega) \,{\rm e}^{{\rm i}(\kappa\rho-\omega\tau)}
  \,{\rm d}\kappa\,{\rm d}\omega.
\end{eqnarray*}
The wavenumber spectrum $\Theta(\kappa)$ and the frequency spectrum
$\Phi(\omega)$ are both related to $\Psi$ via
\begin{equation}
\label{eq:time_and_space_spectra}
\Theta(\kappa) = \int_{-\infty}^\infty \Psi(\kappa,\omega) \,{\rm d}\omega,
\quad
\Phi(\omega) = \int_{-\infty}^\infty \Psi(\kappa,\omega) \,{\rm d}\kappa,
\end{equation}
whence the mean-square of $u$ fluctuations can be recovered from
\begin{equation}
\langle u'^2 \rangle =
 \int_{-\infty}^\infty \int_{-\infty}^{\infty}
 \Psi(\kappa,\omega)
  \,{\rm d}\kappa\,{\rm d}\omega \nonumber\\
  = \int_{-\infty}^\infty \Phi(\omega)\,{\rm d}\omega
  =\int_{-\infty}^\infty \Theta(\kappa)\,{\rm d}\kappa.
\end{equation}
These are even, $\Phi(\omega) = \Phi(-\omega)$
and $\Theta(\kappa) = \Theta(-\kappa)$.
Since $\Phi(\omega)$ can be measured directly using hot wires,
the one-dimensional longitudinal spectrum,
\begin{equation}
\label{eq:spectrum_map}
\Theta(\kappa)\equiv\omega'(\kappa)\Phi(\omega(\kappa)),
\end{equation}
strictly defined in terms of a dispersion relation $\omega(\kappa)$
and associated group velocity $\omega'(\kappa) \equiv \mathrm{d}\omega/
 \mathrm{d}\kappa$, is often
approximated using Taylor's frozen-turbulence hypothesis.  Under this assumption, the dispersion relation can be formally written as $\omega_T(\kappa) = U \kappa$ and,
from (\ref{eq:spectrum_map}), $\Theta(\kappa)=U\Phi(U \kappa)$.
%\bjm{Perhaps add that the Fourier spectrum lends itself to a wave-like interpretation?}
This is another interpretation to Taylor's hypothesis based on the Fourier decomposition: the
group velocity $\omega_T'(\kappa)=U$ and phase velocity
$\omega_T/\kappa=U$ of all eddies contributing to the turbulent kinetic energy at a particular wall-normal location
are constants independent of wavenumber and equal to the mean velocity at that location, an approximation accurate for sufficiently small eddies.
To obtain $\Theta$ from $\Phi$, this strict interpretation of
Taylor's hypothesis can be relaxed to account for energetic eddies which travel at $U\pm\Delta U$ provided $\Psi(\kappa, \omega)$
is symmetric with respect to the $\omega = U \kappa$ line, that is
\begin{equation}
\label{eq:symmetric_spectrum}
\Psi(\kappa, \omega) = \Psi(\omega/U, U \kappa),
\end{equation}
because (\ref{eq:symmetric_spectrum}) then relates the two definitions in
(\ref{eq:time_and_space_spectra}):
\[
\Theta(\kappa) \equiv
  \int_{-\infty}^{\infty} \Psi(\kappa, \omega)\,\mathrm{d}\omega
   =  \int_{-\infty}^{\infty} \Psi(\omega/U,U\kappa)\,\mathrm{d}\omega
   =  U\int_{-\infty}^{\infty} \Psi(\kappa',U\kappa)\,\mathrm{d}\kappa'
   \equiv U\Phi(U \kappa).
\]
For the purpose of obtaining wavenumber spectrum from frequency spectrum,
a test of the validity of Taylor's hypothesis
can be recast as a question of the symmetry of
$\Psi(\kappa,\omega)$ with respect to the line $\omega = U \kappa$.
When this symmetry is broken, a different dispersion relation
$\omega_c(\kappa)$ is necessary to relate $\Phi$ to $\Theta$.
Since we have access to both $\Phi$ and $\Theta$, we compute the
$\omega_c(\kappa)$ directly by finding the inverse to
the monotonic function $K_t$
\[
\omega_c(\kappa) = K_t^{-1}(K_x(\kappa)), \quad
K_x(\kappa) \equiv \int_{\kappa}^{\infty}\Theta(\kappa')\,\mathrm{d}\kappa',
\quad
K_t(\omega) \equiv \int_{\omega}^{\infty}\Phi(\omega')\,\mathrm{d}\omega',
\]
where $K_x(0) = K_t(0) = \langle u'^2\rangle/2$. The analogy to the convection velocity of individual eddies is complex.  To first order, $\omega_c(\kappa)$ describes the apparent passing frequency of the most energetic eddies with wavenumber $\kappa$, although this is an integral effect over the range of energetic spanwise wavenumbers.

When computing $\Psi$ from the present LES simulation, a normalised Hann
window in time,
$\sqrt{2/3}\left[1-\cos(2\upi n_t/M_t)\right]$
($M_t\Delta_t$ is the temporal window size),
is applied to $\overline{u}$ before taking the discrete Fourier transform
because the velocity is not periodic in time.
From run G1b (table \ref{tab:runs2}) where $N_t = 22000$ and $M_t = 576$,
the spectrum is averaged across
$\lfloor 22000/(576/2)-1 \rfloor = 75$ half-overlapping windows.
No windowing is necessary in the periodic streamwise direction.

\begin{figure}
\centerline{\includegraphics{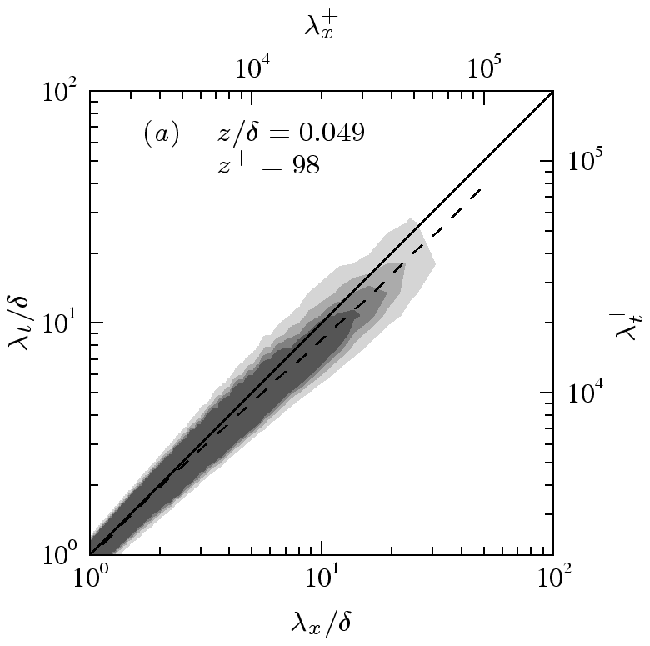}
            \includegraphics{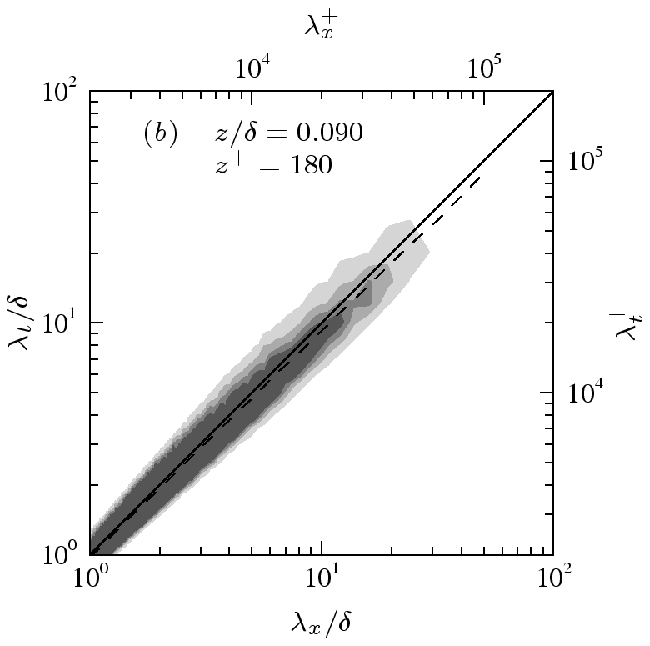}}
\centerline{\includegraphics{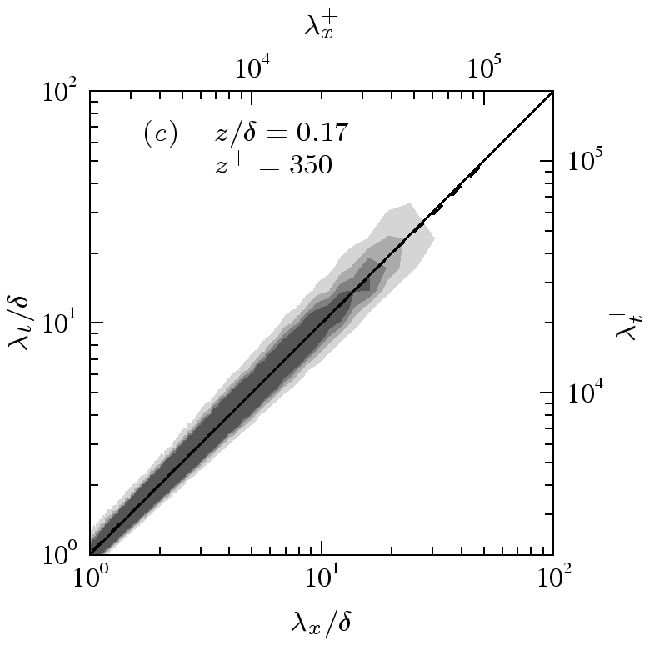}
            \includegraphics{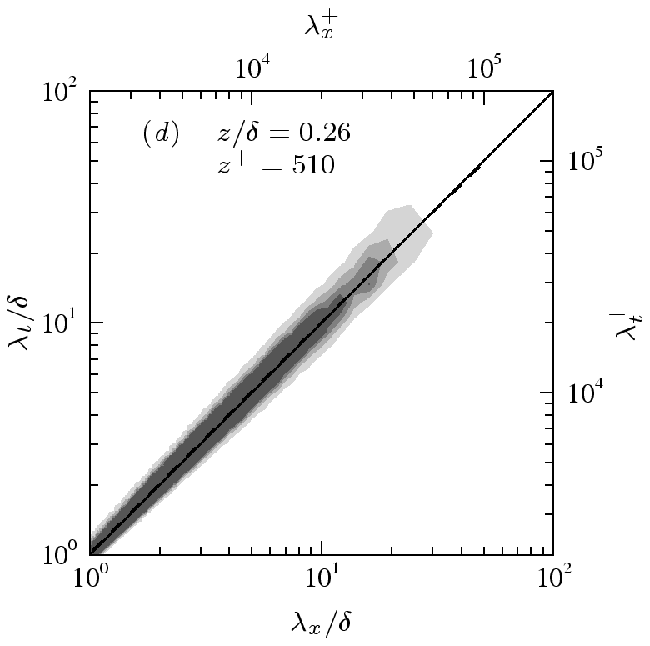}}
\centerline{\includegraphics{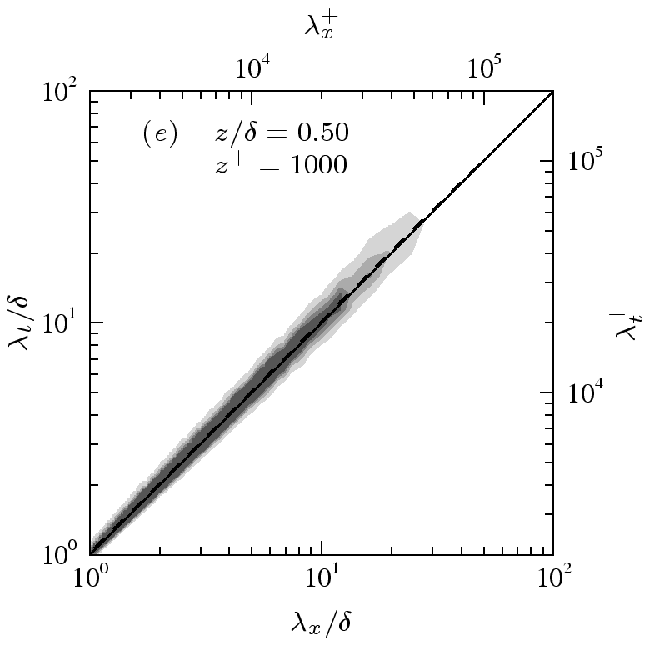}
            \includegraphics{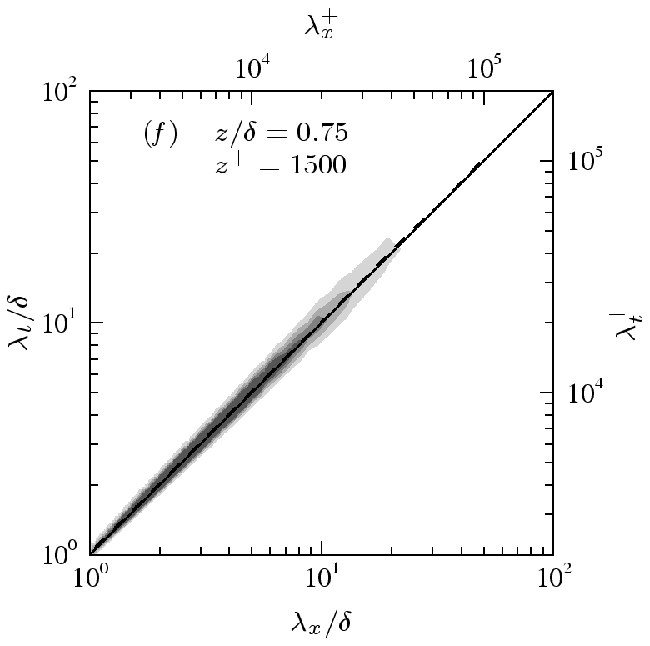}}
\caption{Premultiplied spatio-temporal spectra of streamwise velocity
         fluctuations,
         $\kappa_x \kappa_t \Psi/u_\tau^2 = 0.1, 0.2, 0.3, 0.4$,
         at various heights from $\Rey_\tau = 2\,{\rm k}$ channel flow LES,
         run G1b (table \ref{tab:runs2}):
         \dashed, dispersion relation computed from $\Psi$,
         $\lambda_t = \lambda_x \omega/\omega_c$; \solid,
         Taylor's hypothesis $\lambda_t = \lambda_x$.}
\label{fig:xt_spectra}
\end{figure}
We plot in figure \ref{fig:xt_spectra} contours of the premultiplied spectrum,
$\kappa_x \kappa_t \Psi/u_\tau^2$ versus
$\log\lambda_x$ and $\log\lambda_t$, where $\lambda_x = 2\upi/\kappa_x$,
$\lambda_t = 2\upi/\kappa_t$ and $\kappa_t = \omega/U$.
When Taylor's hypothesis is valid, contours of
$\kappa_x \kappa_t \Psi/u_\tau^2$
should be symmetrical about the $\lambda_t=\lambda_x$ line.
Observe from figure \ref{fig:xt_spectra} that Taylor's hypothesis
is indeed a good approximation, except near the wall,
$z/\delta=0.041$ and $0.083$, and for the large scales,
$\lambda_t/\delta$, $\lambda_x/\delta > 10$.   Note that the LES formulation
does not permit effective examination of smaller scales or locations closer to the wall.
The dispersion relation line computed from $\Psi$,
$\lambda_t = \lambda_x \omega/\omega_c$,
appearing below (to the right) of Taylor's hypothesis,
$\lambda_t = \lambda_x \omega/\omega_T = \lambda_x \cdot 1$, implies that
the phase velocity
$U_c \equiv \omega_c/\kappa$ is larger than the mean velocity $U$.

Despite overwhelming similarities, the phase velocity $U_c$
is not strictly the same as the convection velocity defined by
(2.7) from \cite{Wills1964}, where it is related to the ridge of $\Psi$.
Recall from (\ref{eq:spectrum_map}) that by construction
$U_c(\kappa)$ is defined such that
the wavenumber spectrum $\Theta$ can be recovered from the frequency spectrum
$\Phi$, that is
$\Theta(\kappa) \equiv \Phi(\kappa U_c(\kappa))
\mathrm{d}(\kappa U_c(\kappa))/\mathrm{d}\kappa$.
Put another way, it is the mapping from wavenumber space to frequency space
such that the energy observed in wavenumber space at the wavenumber
$\kappa$, $\Theta(\kappa)\,\mathrm{d}\kappa$, is equal to
that observed in frequency space $\Phi(\kappa U_c(\kappa))
\mathrm{d}(\kappa U_c(\kappa))$ at frequency
$\omega = \kappa U_c(\kappa)$ (see also discussion in \cite{Monty2009}).
In practice, see figure \ref{fig:xt_spectra},
this integral-based definition often traces out the ridge of
$\Psi$, that is the definition in \cite{Wills1964}.
Because it can be measured readily, the ridge-based definition
is often taken as the surrogate for $U_c$.
Energy transmission occurs at neither $U_c$ nor the velocity given by
\cite{Wills1964}, but
the group velocity $\mathrm{d}(\kappa U_c(\kappa))/\mathrm{d}\kappa$,
that is the velocity of the energy of wavepackets with wavenumber
near $\kappa$.
Summarising, to obtain the wavenumber spectrum from the frequency spectrum,
we require both the phase velocity and group velocity, see
(\ref{eq:spectrum_map}): the former appearing in the argument
of $\Phi$ as $\kappa U_c(\kappa)$
to rescale the frequency; and
the latter appearing as a factor of $\Phi$ as
$\mathrm{d}(\kappa U_c(\kappa))/\mathrm{d}\kappa$ to rescale the
rate of change in frequency.

\begin{figure}
\centerline{\includegraphics{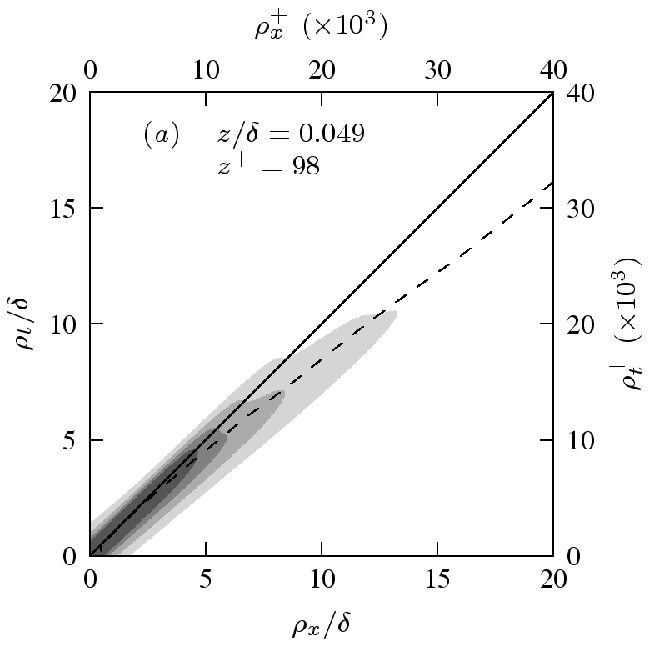}
            \includegraphics{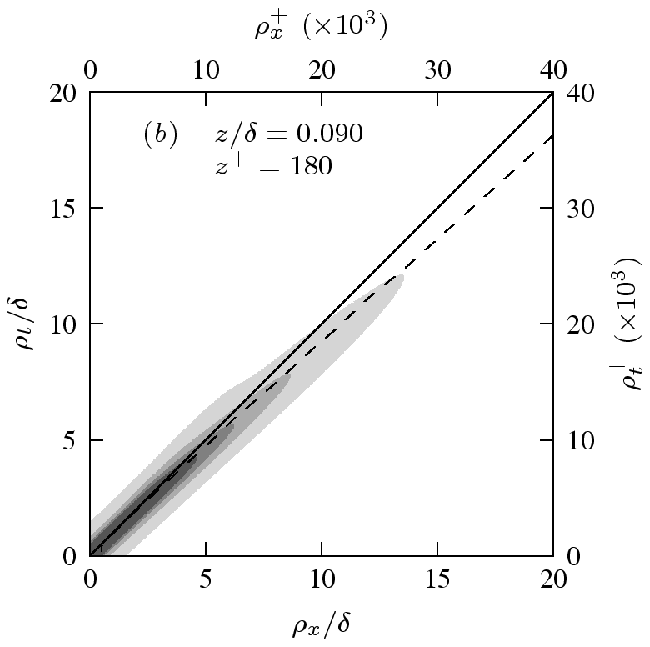}}
\centerline{\includegraphics{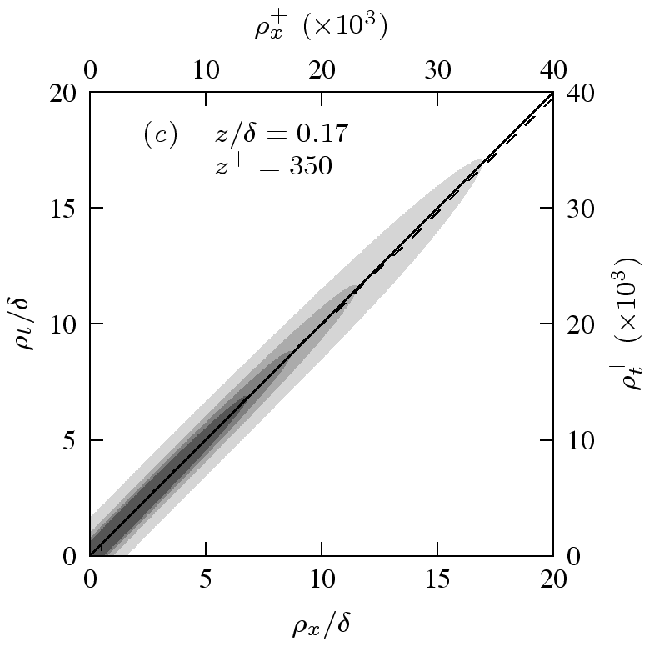}
            \includegraphics{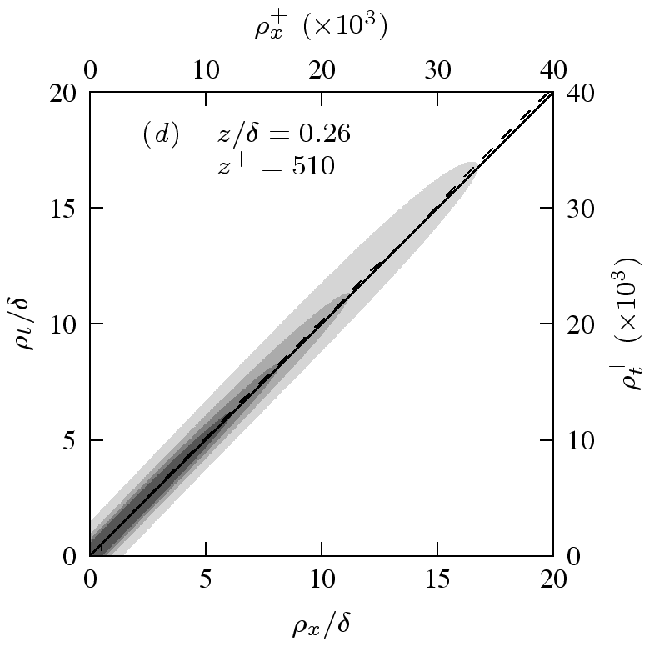}}
\centerline{\includegraphics{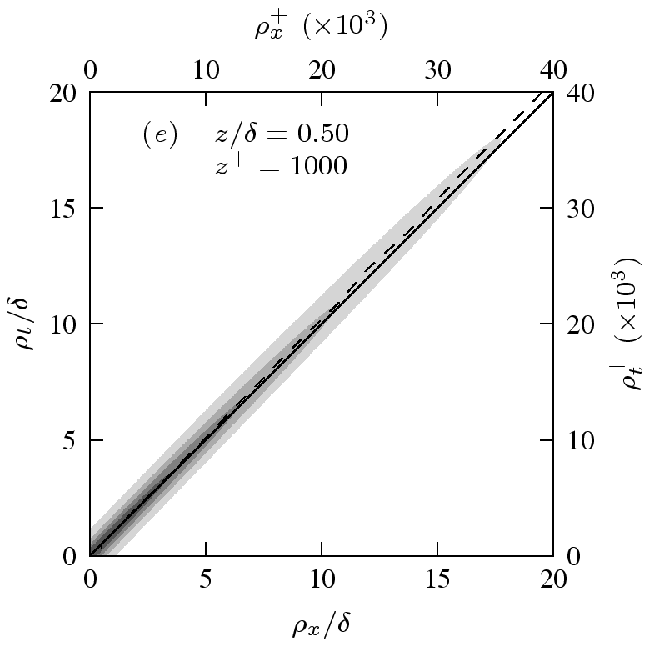}
            \includegraphics{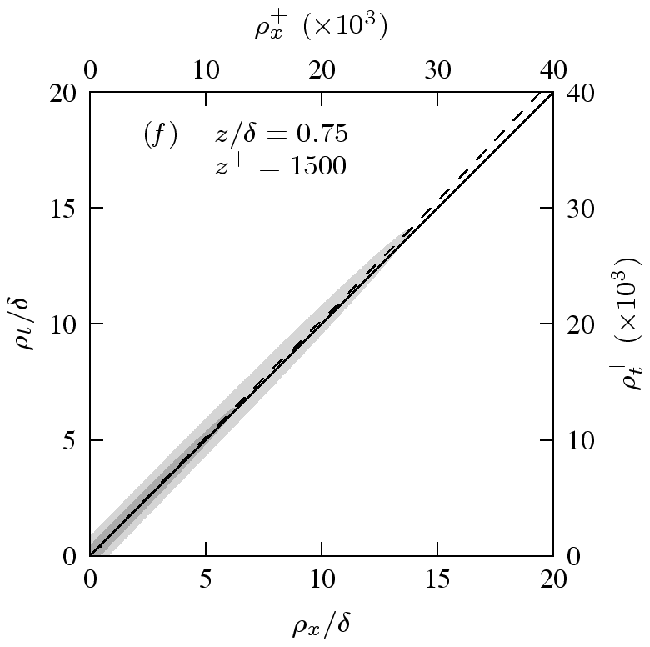}}
\caption{Spatio-temporal correlations of streamwise velocity fluctuations,
         $R/u_\tau^2 = 0.3, 0.6, 0.9, 1.2$, at various heights
         from $\Rey_\tau = 2\,{\rm k}$ channel flow LES,
         run G1b (table \ref{tab:runs2}):
         \dashed, dispersion relation computed from $\Psi$,
         $\rho_t = \rho_x\omega/\omega_c$; \solid,
       Taylor's hypothesis $\rho_t = \rho_x$.}
\label{fig:xt_correlation}
\end{figure}

Following \cite{Dennis2008}, we can also test the validity of
Taylor's hypothesis in physical space
by plotting the autocorrelation $R$ defined by
(\ref{eq:autocorrelation}).
Taylor's hypothesis is valid, or more precisely there is a straightforward conversion from the temporal to the spatial domain, where the contours of $R$ are
symmetrical about the $\rho_x = \rho_t$ line, where $\rho_t = \tau U$.
Like figure \ref{fig:xt_spectra}, $R$ in
figure \ref{fig:xt_correlation}
shows an unequivocal departure from Taylor's hypothesis near the wall,
$z/\delta=0.041$ and $0.083$
and for the large scales, $\rho_t/\delta$, $\rho_x/\delta > 10$.
There also appears to be a slight discrepancy of the opposite sign for the large scales at $z/\delta= 0.5$, which is more marked in the autocorrelation than our presentation of the spectrum.
The boundary layer PIV experiment performed by \cite{Dennis2008}
at $Re_\theta \approx 4.7\,{\rm k}$ reported that
Taylor's hypothesis is still valid at the height
$z/\delta = 0.16$ for the field of view
$\rho/\delta< 0.29\,{\rm m}/0.09\,{\rm m} = 3.2$
and $\tau U/\delta < 1\,{\rm s}
\times 0.57\,{\rm m}{\rm s}^{-1}/0.09\,{\rm m} = 6.3$.
This is consistent with the present LES data since
at $z/\delta=0.17$, figure \ref{fig:xt_correlation}({\it c}),
$R$ is indeed symmetrical about the $\rho_x = \rho_t$ line,
even up to very large scales $\rho_t$, $\rho_x = 20\delta$.

Consideration of spatio-temporal spectra permits some speculation about the convection velocities of the energetic structures and the error associated with identifying the footprint of eddies with a particular streamwise scale on the near-wall region from temporal data. The LES velocity fields unequivocally indicate that the most energetic large structures convect faster than the local mean velocity close to the wall, with the deviation growing close to the wall.
Conversely, these large structures convect slower than
the local mean velocity near the channel centre, figure \ref{fig:xt_correlation}(\textit{e}, \textit{f}).
This suggests that these eddies are ``local'' to a region in the overlap layer, in the sense that the mean velocity matches their convective velocity somewhere in the log region.
We speculate that the location of this velocity matching corresponds to the location of the large-scale streamwise energy peak, which is consistent with the approximate magnitude of the difference between $\omega/\omega_T$ and $\omega/\omega_c$ for large wavelength.
%\bjm{Top of the large-scale structure showing convection velocity slower than the mean flow -- conditional averages certainly show that the dominant structure reaches far from the wall.}
This suggests that the departure from Taylor's hypothesis at the large scales should strengthen with increasing Reynolds number due to the increasing shear near the wall.  This is an area of current experimental study.

\subsection{Large-scale--small-scale interaction}\label{sect:R}

With the differences between the spatial and the temporal decompositions in mind, we now describe the correlation that characterizes the interaction between the large-scales and the small-scales.
We first extract the large-scale fluctuations $u_L$ by applying a sliding-window
top-hat time average, centred at $t$, to $u$:
\begin{equation}
\label{eq:large_scale}
u_L(t) =
\dfrac{1}{\tau}\int_{t-\tau/2}^{t+\tau/2} u(t')\,{\rm d}t',
\end{equation}
where $\tau$ is the width of the sliding window.
A low-pass filter, (\ref{eq:large_scale}), dampens fluctuations with
frequencies higher than $1/\tau$.
In spectral space, (\ref{eq:large_scale}) is equivalent to
a multiplication by the filter $\sin(\tau \omega/2)/(\omega/2)$,
where $\omega$ is the angular frequency.
For clarity, we have suppressed the ${\bm x}$-dependence of
$u$ in this part of the discussion since ${\bm x}$ is held constant.

The small-scale fluctuations are defined to be the remaining part of the motion,
$u_S=u-u_L$.
The small-scale intensity can be measured by its local r.m.s.,
\begin{equation}
\label{eq:small_scale}
\widetilde{u}_S(t) = \left(
\dfrac{1}{\tau}\int_{t-\tau/2}^{t+\tau/2} u_S^2(t')\,{\rm d}t'
\right)^{1/2}.
\end{equation}
Physically, $\widetilde{u}_S$ measures the local envelope or intensity of
small-scale fluctuations.
For example, if $u_S$ is normally distributed, $95\%$ of its amplitude
is estimated to lie within $2 \widetilde{u}_S$.
If $u_S$ is not normal, $\widetilde{u}_S$ still measures the spread or
envelope of $u_S$.
In any case, $\widetilde{u}_S^2$ appears in the equations governing
$u_L$, obtained by applying the filter (\ref{eq:large_scale})
to the Navier--Stokes equations (see \cite{Reynolds1972} for a related two-scale decomposition),
which is another way to interpret $\widetilde{u}_S$. Note that equivalent approaches have been used by \cite{Bandyopadhyay1984} and \cite*{Guala09mod} to obtain similar results in a range of flows including a laboratory turbulent boundary layer and the near-wall region of the near neutrally stable atmospheric surface layer, respectively.

An elegant alternative to obtain the envelope of $u_S$
is via the Hilbert transform \cite[]{Mathis2009}, and this approach has led to a significant advance in understanding of the large-small scale interactions.
However it is harder to relate the results to the governing equations of turbulence.

When calculating an r.m.s.\ defined locally, (\ref{eq:small_scale}), one has
to contend with the inevitability that large-scales (low frequencies)
have been aliased into the small-scale signature.
Perhaps a better alternative is to use a tapered window in
(\ref{eq:small_scale}), alleviating some but not all of the aliasing.
We have tried this and found some minor changes, but the general
picture is unaltered, and so
%Considering other approximations inherent in LES
we decided to keep
the simple definition in (\ref{eq:small_scale}).
The Hilbert transform bypasses this aliasing difficulty at
the enveloping stage, but the issue reappears when one filters the
envelope signal.
We note that even if a perfect decomposition can be found, nature herself
does not permit it, that is the two peaks in $\kappa_x E_{uu}$ \cite[]{Hutchins2007a}
are never completely isolated, at least in Fourier space.

In terms of LES quantities,
we can write (\ref{eq:large_scale}) and (\ref{eq:small_scale}) as
\begin{subequations}
\label{eq:sliding_time_average}
\begin{eqnarray}
u_L(t) &=&
\dfrac{1}{\tau}\int_{t-\tau/2}^{t+\tau/2}
  \overline{u}(t')\,{\rm d}t', \\
\widetilde{u}_S(t) &=&
\left(\dfrac{1}{\tau}\int_{t-\tau/2}^{t+\tau/2}
  \left[u_S^2(t') +T_{xx}(t')\right]\, {\rm d}t'\right)^{1/2},
\end{eqnarray}
\end{subequations}
where $\overline{u}$ is the resolved velocity;
$u_S = \overline{u} - u_L$; and
$T_{xx}$ is the modelled subgrid fluctuations associated with time scales
smaller than the
numerical discretization $\Delta_t$.
Using (\ref{eq:sliding_time_average}), we now construct the normalised
large-scale--small-scale correlation based on temporal filtering:
\begin{equation}
\label{eq:large_small_corr}
R_\tau(z)
  = \dfrac{\langle (u_L - U) (\widetilde{u}_S - \langle \widetilde{u}_S \rangle) \rangle}
    {\langle(u_L - U)^2\rangle^{1/2}
     \langle (\widetilde{u}_S - \langle \widetilde{u}_S \rangle )^2\rangle^{1/2}},
\end{equation}
where the global or ensemble average is formally given by
\begin{equation}
\label{eq:global_average}
\langle \phi \rangle \equiv \lim_{T\rightarrow \infty}\dfrac{1}{T}
  \int_{-T/2}^{T/2}\phi(t')\,\mathrm{d}t'.
\end{equation}
Note that the visual envelope of $u_S$, e.g.\ $2\widetilde{u}_S$,
does not affect $R_\tau$ because the constant factor cancels out
in the normalised correlation (\ref{eq:large_small_corr}).
In other words, $R_\tau$ does not contain amplitude information;
it does, however, contain phase information, since it is the cosine of the angle (or phase)
between $u_L-U$ and $\widetilde{u}_S-\langle\widetilde{u}_S\rangle$,
using the inner product $\langle \; \rangle$.
In practice, we obtain $R_\tau(z)$ by
replacing the integrals with sums and ensuring the recording period
$T$ is much larger than the largest physical time scale in the flow, see
table \ref{tab:runs2}.

The spatial counterpart to (\ref{eq:large_small_corr}), $R_\rho(z)$,
is defined analogously, with $x$ and $\rho$ respectively replacing $t$ and $\tau$,
while holding other variables constant.
For $R_\rho$, the global average (\ref{eq:global_average}) is replaced by an
average over the wall-parallel plane with area $L_x L_y$.
The spatial correlations are calculated from runs G1 and H3 at one
snapshot in time.

The physical meaning of the correlations $R=R_\tau$, $R_\rho$
are as follows.
If large-scale higher-speed regions carry higher small-scale intensity (positively correlated, in phase),
then $R \approx 1$.
Similarly, if large-scale higher-speed regions carry lower
small-scale intensity (negatively correlated, $\upi$ out of phase), then $R\,\approx -1$. $R \,\approx0$ can occur either if there is no correlation between the large and small scales, or if they are $\upi/2$ out of phase, which is physically the more likely option given the strong correlation for small and large $z/\delta$, as already demonstrated by \cite{Bandyopadhyay1984, Mathis2009}.

Although the r.m.s.-based correlation coefficient
(\ref{eq:large_small_corr}) is different from its Hilbert-transform-based
counterpart in the boundary layer study of \cite{Mathis2009}, we expect similar
qualitative features if the large-scale--small-scale phase relationship
is a universal aspect
of wall-bounded flows, namely channels and boundary layers.
As we are interested in large scales with sizes of the order $10\delta$,
we expect that the LES will capture this statistic satisfactorily
since the premise of an LES is to directly simulate the large scales;
presently, there are $6$ grid points per $\delta$ in both wall-parallel
directions and $24$ grid points per $\delta$ in the wall-normal direction.

\begin{figure}
\centerline{\includegraphics{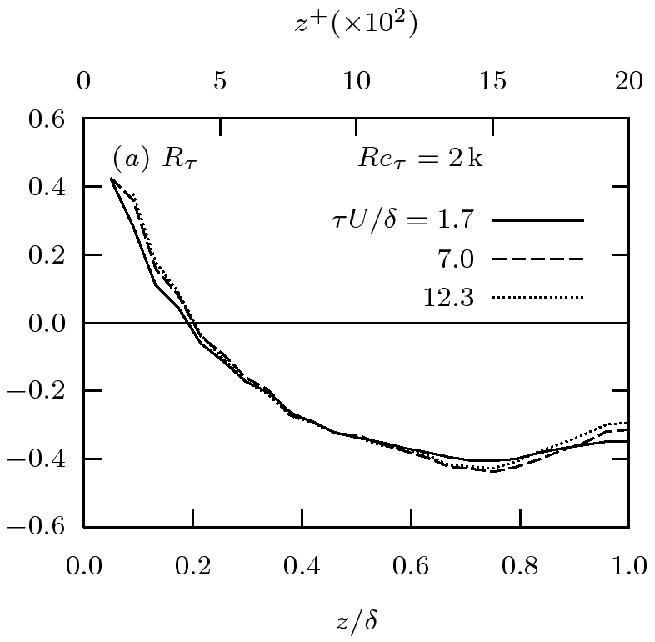}
            \includegraphics{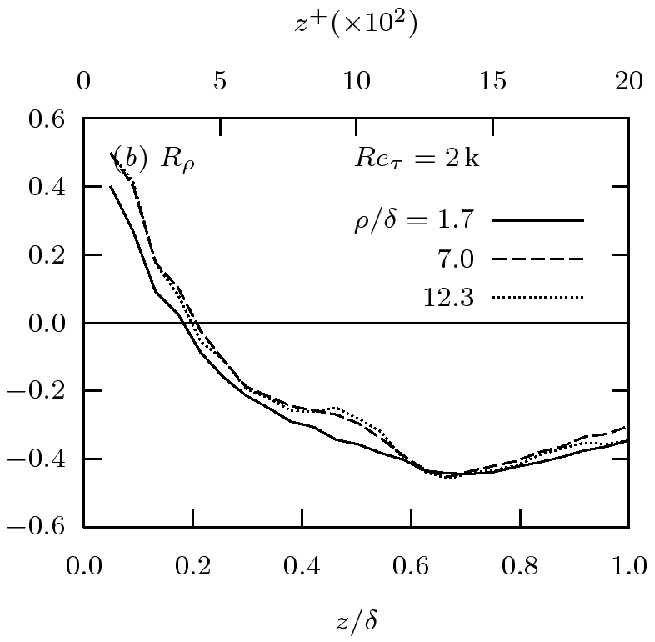}}
\centerline{\includegraphics{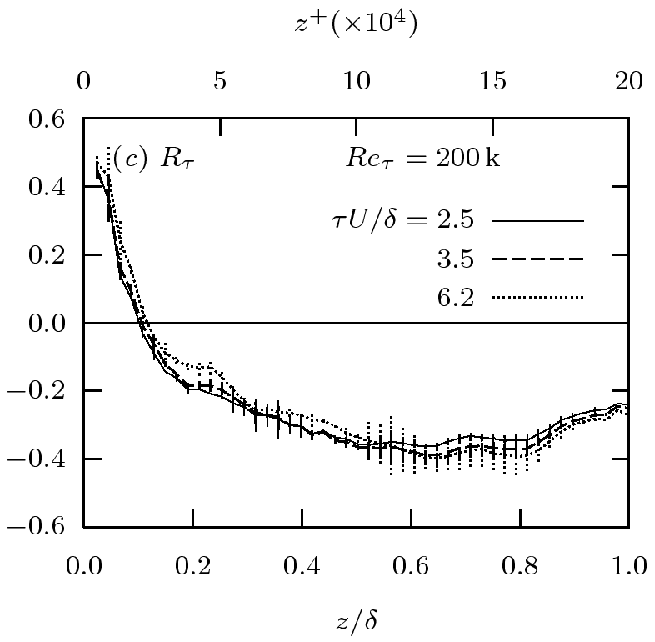}
            \includegraphics{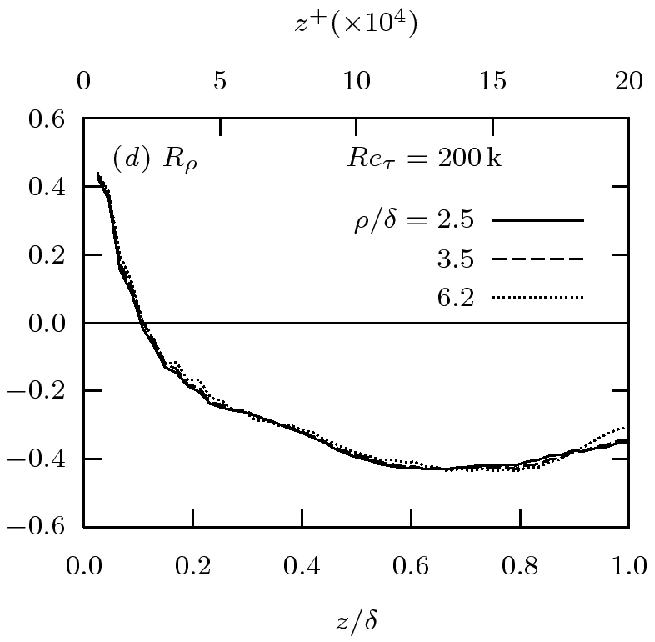}}
\caption{Profiles of large-scale--small-scale correlations,
         $R_\tau$ and $R_\rho$, defined by (\ref{eq:large_small_corr}).
         Filter sizes and $\Rey_\tau$ inset.
         Error lines in (\textit{c}) indicate convergence uncertainties
         from inadequate averaging period $T$.
         Data from channel flow LES, runs G1 and H3 (table \ref{tab:runs2}).}
\label{fig:corr_profile}
\end{figure}
Figure \ref{fig:corr_profile} compares the correlations
based on temporal filtering, $R_\tau$, and correlations based on spatial
filtering, $R_\rho$, for
$\Rey_\tau=2\,{\rm k}$ and $\Rey_\tau=200\,{\rm k}$ and different values of $\tau$ and $\rho$.
Observe that near the wall, $u_L$ and $\widetilde{u}_S$ are
positively correlated, up to $R \approx 0.4$ (the maximum in the domain we resolve, although note that the maximum value likely increases closer to the wall), but above
a certain crossing height, $z/\delta \approx 0.2$ for $\Rey_\tau = 2\,{\rm k}$
and $z/\delta \approx 0.11$ for $\Rey_\tau = 200\,{\rm k}$,
they are negatively correlated, down to $R \approx -0.4$.
The trend of decreasing crossing height with increasing Reynolds number
is also reported by \cite{Mathis2009} for the turbulent boundary layer, with $z/\delta \approx 0.07$ for
$\Rey_\tau = 2.8\,{\rm k}$ and $z/\delta \approx 0.03$ for
$\Rey_\tau = 19\,{\rm k}$.
Presently, the correlations are largely independent of filter sizes
$1.7 < \tau U/\delta, \rho/\delta < 12.3$, although the deviation between $R_\rho$ and $R_\tau$ with the larger filter sizes from the one with $\rho/\delta = 1.7$ close to the wall is exacerbated in the spatial plots, as would be expected from the arguments concerning Taylor's hypothesis at the large scales in the preceding section.
%The reason for the kick up of the curves corresponding to $\rho/\delta = 7.0, 12.3$ in the vicinity of $z/\delta = 0.45$ is not known, with no obvious numerical or structural origin.
The kick-up of $R_\rho$ relative to $R_\tau$ for curves corresponding to
$\rho/\delta = 7.0$, $12.3$ in the vicinity of $z/\delta = 0.45$, see
figure \ref{fig:corr_profile}(\textit{a}, \textit{b}) is presumably related to
the convection velocity effect demonstrated in
figure \ref{fig:xt_correlation}(\textit{e}).
%INDIFFERENCE TO FILTER SIZE IN TIME MEANS ITS AT LARGE FREQUENCY}
%\bjm{Another interesting feature of figure~\ref{fig:corr_profile} is the difference between the temporal and spatial correlations at $\Rey_\tau = 2{\rm k}$ and $z/\delta \sim 0.5$ for the larger filter sizes... May be due to a different structure - Mathis shows that changing the convection velocity changes only the near-wall shape of $R_\tau$. Not observed at the higher Re - convergence issues.}
A small sensitivity to filter sizes is reported by \cite{Mathis2009}.
Perhaps a precise quantitative comparison is impossible owing to the different
envelope-extraction techniques and the different type of wall-bounded flows.
Note that a robust feature is that $R$ increases slightly at the centre of
the channel, but remains negative (figure \ref{fig:corr_profile}).
This increase is also reported in \cite{Mathis2009} (and this
sensitivity appears to be slightly enhanced in the case of the spatial correlation), but their increase is
from negative correlations to positive correlations, a feature possibly
related to the intermittent boundary layer thickness not present in channel flows.

Although the large-scale--small-scale interaction was recently \cite[]{Mathis2009} %Figure~\ref{fig:corr_profile} has been
framed in terms of amplitude modulation, the results could also be discussed in terms of the relative phase between the large and small scales, as originally posed by \cite{Bandyopadhyay1984} and implied by the formulation of (\ref{eq:large_small_corr}).  %Zero values of $R$ can be obtained either through a negligible amplitude modulation or by a spatial average phase shift between the large and small scale signals of $\upi/2$.  We examine this phase relationship in following section.

\subsection{Conditionally averaged large-scale velocities and small-scale
  intensities}

To gain some insight into the structure of the large-scale coherent regions and the phase relationship %dynamics of the modulation reversal
between $u_L$ and $\widetilde{u}_S$ shown in figure
\ref{fig:corr_profile}, we now turn our attention to
conditionally averaged $u_L$ and $\widetilde{u}_S$ fields computed
with the streamwise filter window $\rho=\delta$.
As seen in figure \ref{fig:corr_profile} and in \cite{Mathis2009},
this phase relationship %modulation
is relatively unaffected by the choice of
$\rho$.

\begin{figure}
\centerline{\includegraphics{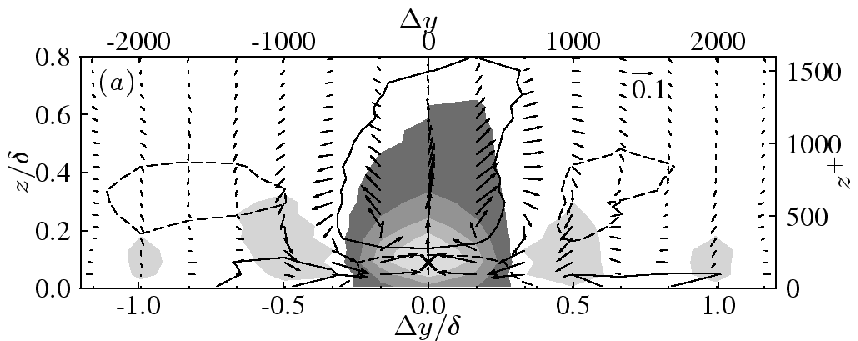}}
\centerline{\includegraphics{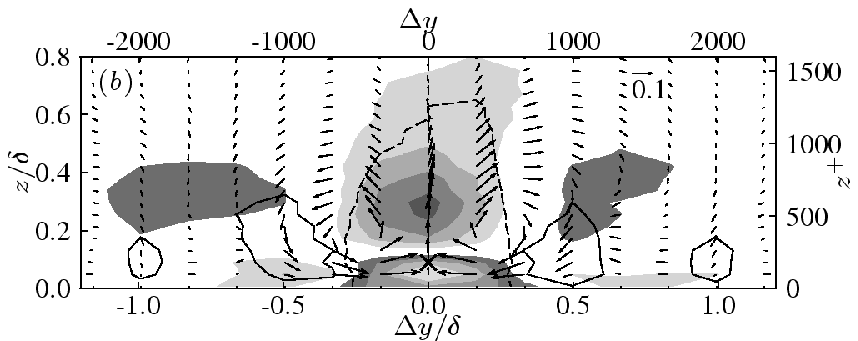}}
\centerline{\includegraphics{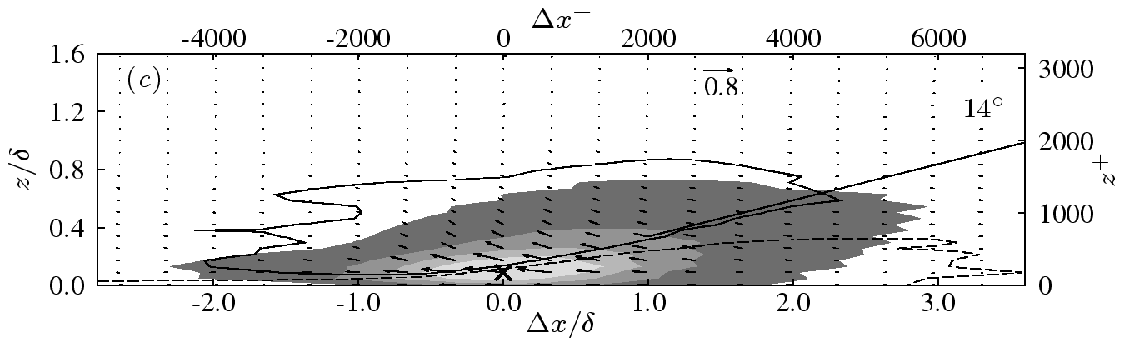}}
\centerline{\includegraphics{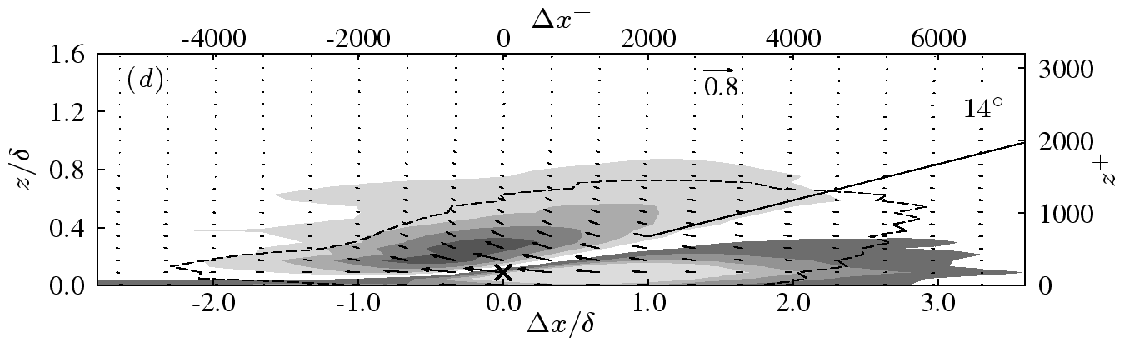}}
\caption{Conditionally averaged streamwise velocity fields in the  (\textit{a}, \textit{b}) spanwise--wall-normal and  (\textit{c}, \textit{d}) streamwise--wall-normal planes, with
         $\langle {\bm u}_L^+ | A_1\rangle(\Delta {\bm x})$ and
         $\langle \widetilde{u}_S^+|A_1\rangle(\Delta {\bm x})$,
         where $A_1$ is the low-speed event $u_L({\bm x}_\times) - U < 0$
         with probability $P\{A_1\}=0.53$ at
         ${\bm x}_\times/\delta = (0, 0, 0.090)$:
         (\textit{a}, \textit{c}) filled contours of
         $|\langle u_{L}^+|A_1\rangle - U^+| = 0.1, 0.3, 0.5, 0.7$,
         line contours of
         $|\langle \widetilde{u}_S^+|A_1\rangle
           - \langle\widetilde{u}_S^+\rangle| = 0.01$; and
         (\textit{b}, \textit{d}) filled contours of
         $|\langle \widetilde{u}_S^+|A_1\rangle
          - \langle\widetilde{u}_S^+\rangle|
         = 0.01, 0.02, 0.03, 0.04$,
         line contours of
         $|\langle u_L^+ |A_1\rangle - U^+| = 0.1$.
         Darker to lighter shades and dashed lines for negative values.
         Vectors represent in-plane velocity components.
         Data from snapshot of $\Rey_\tau = 2\,{\rm k}$ channel flow LES,
         run G1b (table \ref{tab:runs2}).}
\label{fig:conavg_negu}
\end{figure}

Figure \ref{fig:conavg_negu} shows ensemble averages,
$\langle u_L | A_1\rangle (\Delta {\bm x})$ and
$\langle \widetilde{u}_S | A_1 \rangle (\Delta {\bm x})$, conditioned on
the occurrence of a large-scale low-speed event at $z/\delta = 0.090$,
$A_1 = u_L({\bm x}_\times) - U < 0$,
where $\Delta {\bm x} = {\bm x} - {\bm x}_\times$,
${\bm x}_\times^+ = (0, 0, 180) \Leftrightarrow
{\bm x}_\times/\delta = (0, 0, 0.090)$, computed from a snapshot of LES run G1b
(table \ref{tab:runs2}).
The spanwise--wall-normal view, figure \ref{fig:conavg_negu}(\textit{a}),
is previously shown in figure 7(\textit{a}) of \cite{Hutchins2007a} with the
choice $z_\times^+ = 150$ from the DNS
data of \cite{delAlamo2004} at $\Rey_\tau \approx 1 {\rm k}$.
For reference,
the vectors in figure \ref{fig:conavg_negu}(\textit{a}, \textit{b})
correspond to LES discretization points. The agreement between DNS and LES results is striking. The essential features of the DNS averages are also seen in the
present LES averages, namely 1)
a splatted low-speed region with minimum $\langle u_L^+|A_1\rangle \approx -0.8$ and width
$\Delta y/\delta \approx 0.4$ centred on ${\bm x}_\times$ flanked
on both sides by weaker high-speed regions with maximum $u_L^+ \approx 0.1$;
and 2) the in-plane large-scale swirl at
$\Delta_x/\delta \approx \pm 0.2$, $z/\delta \approx 0.2$. 

We report that $P\{A_1\}=0.53$, implying that nearly identical figures,
but with signs reversed, are seen when we condition on the
large-scale high-speed event $A_1' = u_{L\times} - U > 0$ (the
complement of $A_1$) because
\[0 = \langle \phi \rangle \equiv
    \langle \phi|A_1\rangle P\{A_1\}
  + \langle \phi|A_1'\rangle P\{A_1'\} \quad
\Rightarrow \quad \langle \phi |A_1\rangle \approx -\langle \phi|A_1'\rangle,
\]
(equality holds if $P\{A_1\}$ is exactly $1/2$).
Since the reversed picture, conditioned on $A_1'$ exists,
neighbouring high-speed regions in figure \ref{fig:conavg_negu}(\textit{a})
could be interpreted as equal-magnitude high-speed regions
whose strengths have been smeared by other less-dominant
large-scale dispersive motions in the averaging process.
Thus, a detailed description of physical processes far away from
${\bm x}_\times$ is difficult to ascertain.
One may be tempted to believe from the conditional average fields that
these structures are aligned in the streamwise direction;
instantaneous visualisations, see \cite{Monty2007}, suggest that
these are in fact meandering structures, which would have been
obscured in the averaging over the periodic domain in our study.  However this apparent meandering coherence could equally well be interpreted as adjoined regions of shorter coherence, which are individually well-captured by the conditional averaging. The spanwise scale appears to be approximately $\delta$.

Figure \ref{fig:conavg_negu}(\textit{b}) shows the spanwise structure of the
relationship between $u_L$ and $\widetilde{u}_S$:
near the wall, $u_L$ and $\widetilde{u}_S$ are positively correlated,
but above $z/\delta \approx 0.1$, they are negatively correlated.
This crossing point is different from $z/\delta \approx 0.2$
seen in figure \ref{fig:corr_profile}(\textit{a}).
The discrepancy is resolved by noting the inclusive
($A_1 = u_L - U > 0$) and non-collocated (two-point) nature of
the conditioning used for the averages $\langle u_L|A_1 \rangle$
and $\langle \widetilde{u}_S | A_1 \rangle$, as well as the single plane rather than integral representation.
In contrast, the correlation $R$, see (\ref{eq:large_small_corr}),
is constructed from the one-point collocated statistic,
$\langle u_L \widetilde{u}_S \rangle \equiv
\int \langle u_L | u_L \rangle
  \langle \widetilde{u}_S | u_L \rangle p(u_L)\,{\rm d}u_L$,
which is not the same as
$\sum_A \langle u_L | A \rangle \langle \widetilde{u}_S | A \rangle P\{A\}$.
The opposite $u_L$--$\widetilde{u}_S$ configuration of
the weaker flanking regions in
figure \ref{fig:conavg_negu}(\textit{b}) suggests that
these too experience %modulation
phase reversal.

\begin{figure}
\centerline{\includegraphics{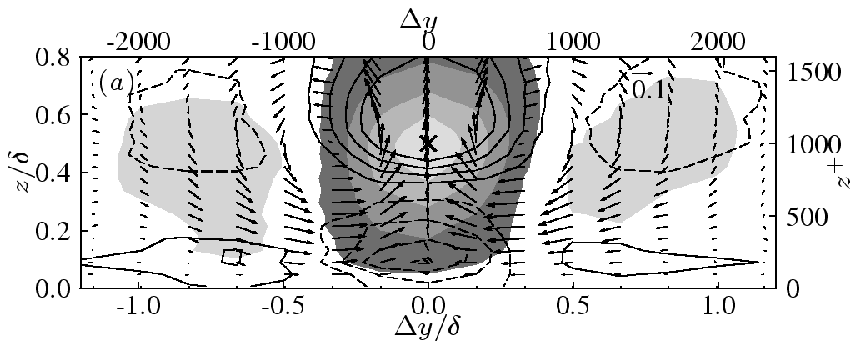}}
\centerline{\includegraphics{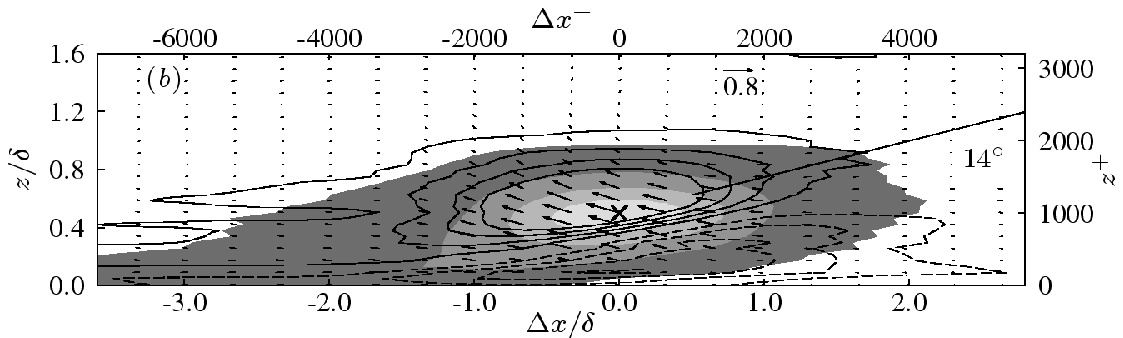}}
\caption{Conditionally averaged streamwise velocity fields in the  (\textit{a}) spanwise--wall-normal and  (\textit{b}) streamwise--wall-normal planes, with
         $\langle {\bm u}_L^+ | A_2\rangle(\Delta {\bm x})$ and
         $\langle \widetilde{u}_S^+|A_2\rangle(\Delta {\bm x})$,
         where $A_2$ is the low-speed event
         $u_L({\bm x}_\times) - U < 0$
         with probability $P\{A_2\}=0.47$ at
         ${\bm x}_\times/\delta = (0, 0, 0.5)$:
         filled contours of $|\langle u_{L}^+|A_2\rangle - U^+| = 0.1, 0.3, 0.5, 0.7$,
         line contours of
         $|\langle\widetilde{u}_S^+|A_2\rangle - \langle\widetilde{u}_S^+\rangle|
          = 0.01, 0.02, 0.03, 0.04$.
         Darker to lighter shades and dashed lines for negative values.
         Vectors represent in-plane velocity components.
         Data from snapshot of $\Rey_\tau = 2\,{\rm k}$ channel flow LES,
         run G1b (table \ref{tab:runs2}).}

\label{fig:conavg_negu1000}
\end{figure}

Figure \ref{fig:conavg_negu}(\textit{c}, \textit{d}) shows
the streamwise--wall-normal structure of the dominant large-scale motion and the relationship between the small and large scales.  Despite a filter size of $\delta$, the coherence indicated in the figure suggests a wavelength of order $6\delta$, suggesting that the very long structures are the dominant contributors to the $u_L({\bm x}_\times) - U < 0$ signal.  Clearly the large-scale coherence has a streamwise phase that is dependent on the wall-normal location, at least where the coherence is strongest.  Close to the wall this phase variation is weak, while the conditional averages with
${\bm x}_\times^+ = (0, 0, 1000) \Leftrightarrow
{\bm x}_\times/\delta = (0, 0, 0.5)$ shown in figure \ref{fig:conavg_negu1000} show that far from the wall the phase variation with increasing $z/\delta$ is also weak, but close to $\upi$ rather than zero. In the intermediate region, the phase changes rapidly with wall-normal distance.  The contours of constant $|\widetilde{u}_S^+ - \langle\widetilde{u}_S^+\rangle|$ in figure \ref{fig:conavg_negu}(\textit{d}) reveal a surface that cuts through the large-scale, low-speed region at a diagonal
such that the region of negative correlation is larger where
$\Delta x < 0$ but smaller where $\Delta x > 0$.
The angle of this separatrix, at least in the aforementioned intermediate region, is about $14^\circ$,
suggesting that the modulation reversal \cite[]{Mathis2009} is related to the structure inclination angle \cite[]{Marusic2007}.

Figure \ref{fig:conavg_negu}(\textit{c}, \textit{d})
and $P\{A_1\}\approx 1/2$ suggest the stylised picture of a
streamwise train of alternating high-speed and low-speed
regions with the shape in figure \ref{fig:conavg_negu}(\textit{c}).
We propose that this sign change of $u_L - U$
determines the shape of the $\widetilde{u}_S$ region over a range of wall-normal distances.
Consider the governing equation for $\widetilde{u}_S^2$, which
contains the production term $-2\widetilde{u}_S^2 \p u_L/\p x$
\cite[]{Reynolds1972}.
Now, $\p u_L/\p x < 0$ in between a high-speed region placed
upstream ($-\Delta x$) of a low-speed region.
Then, the production $-2\widetilde{u}_S^2 \p u_L/\p x > 0$
increases $\widetilde{u}_S^2$, resulting
in the picture \ref{fig:conavg_negu}(\textit{d}).
The opposite mechanism applies in between
a low-speed region placed upstream of
a high-speed region, in which case $-2\widetilde{u}_S^2 \p u_L/\p x < 0$,
a backscatter of small-scale streamwise energy.
This results in the quarter-phase shift between the $u_L$ and $\widetilde{u}_S$
region.

\begin{figure}
\centerline{\includegraphics{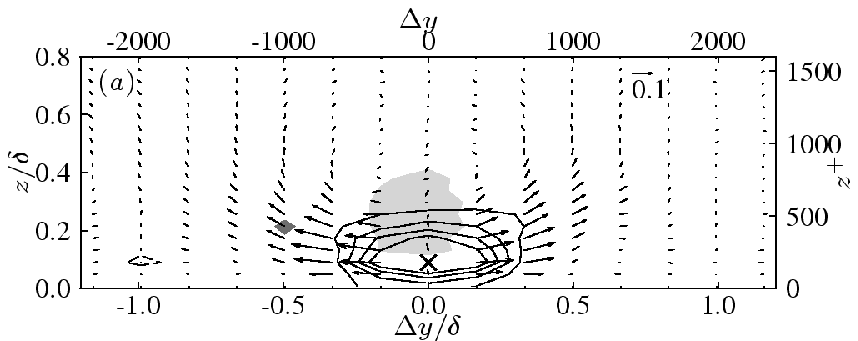}}
\centerline{\includegraphics{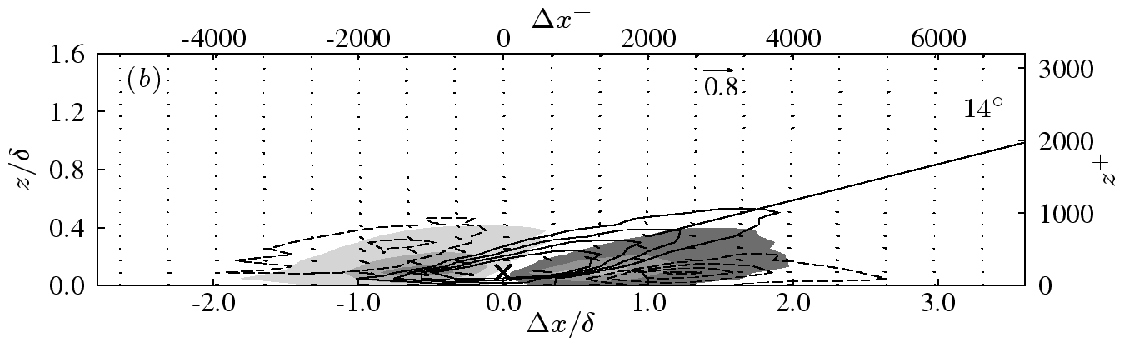}}
\caption{Conditionally averaged streamwise velocity fields in the  (\textit{a}) spanwise--wall-normal and  (\textit{b}) streamwise--wall-normal planes, with
         $\langle {\bm u}_L^+ | A_3\rangle(\Delta {\bm x})$ and
         $\langle \widetilde{u}_S^+|A_3\rangle(\Delta {\bm x})$,
         where $A_3$ is the streamwise high-speed-to-low-speed boundary event $(\p u_L/\p x)({\bm x}_\times) < 0$
         with probability $P\{A_3\}=0.48$ at
         ${\bm x}_\times/\delta = (0, 0, 0.090)$:
         filled contours of
         $|\langle u_{L}^+|A_3\rangle - U^+| = 0.1, 0.3, 0.5, 0.7$,
         line contours of
         $|\langle \widetilde{u}_S^+|A_3\rangle - \langle\widetilde{u}_S^+\rangle| = 0.01, 0.02, 0.03, 0.04$.
         Darker to lighter shades and dashed lines for negative values.
         Vectors represent in-plane velocity components.
         Data from snapshot of $\Rey_\tau = 2\,{\rm k}$ channel flow LES,
         run G1b (table \ref{tab:runs2}).}
\label{fig:conavg_negdudx}
\end{figure}
A clearer picture emerges when we compute averages conditioned on
the $u_L > 0 $ to
$u_L < 0$ boundary, signalled
by the event $A_3 = (\p u_L/\p x)({\bm x}_\times) < 0$
(figure \ref{fig:conavg_negdudx}), interpreted
as a quarter-phase streamwise shift of figure
\ref{fig:conavg_negu}(\textit{c}, \textit{d}).
As expected, $\widetilde{u}_S$ is lowest precisely where
$\p u_L/\p x$ is minimum (at ${\bm x}_\times$),
figure \ref{fig:conavg_negdudx}(\textit{b}).
The bulge-like shape of the $u_L$ regions is preserved, see figure
\ref{fig:conavg_negdudx}(\textit{a}), although with smaller sizes.
Like the average conditioned on $A_1$, this figure is also
reversible, with $P\{A_3\} = 0.48$.

\begin{figure}
\centerline{\includegraphics{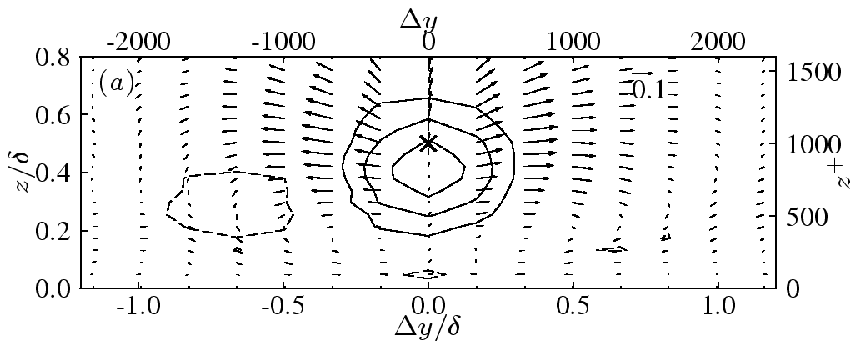}}
\centerline{\includegraphics{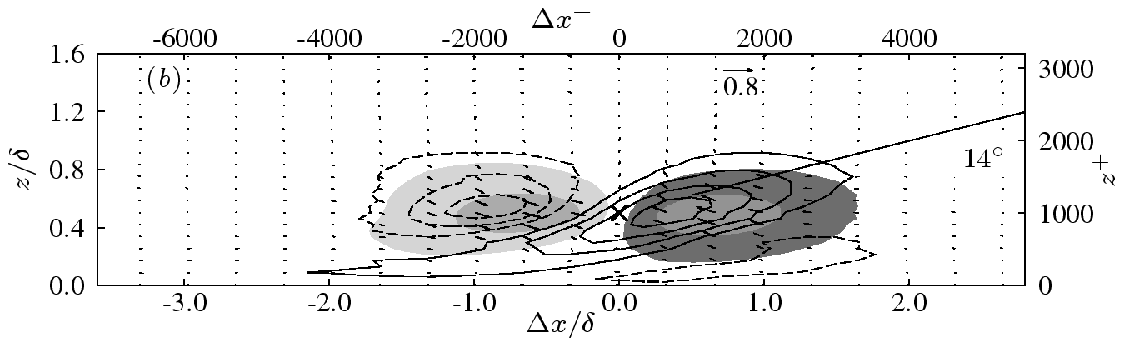}}
\caption{Conditionally averaged streamwise velocity fields in the  (\textit{a}) spanwise--wall-normal and  (\textit{b}) streamwise--wall-normal planes, with
         $\langle {\bm u}_L^+ | A_4\rangle(\Delta {\bm x})$ and
         $\langle \widetilde{u}_S^+|A_4\rangle(\Delta {\bm x})$,
         where $A_4$ is the streamwise high-speed-to-low-speed boundary event
         $(\p u_L/\p x)({\bm x}_\times) < 0$
         with probability $P\{A_4\}=0.48$ at
         ${\bm x}_\times/\delta = (0, 0, 0.5)$:
         filled contours of $|\langle u_{L}^+|A_4\rangle - U^+| = 0.1, 0.3, 0.5, 0.7$,
         line contours of
         $|\langle \widetilde{u}_S^+|A_4\rangle - \langle\widetilde{u}_S^+\rangle|
         = 0.01, 0.02, 0.03, 0.04$.
         Darker to lighter shades and dashed lines for negative values.
         Vectors represent in-plane velocity components.
         Data from snapshot of $\Rey_\tau = 2\,{\rm k}$ channel flow LES,
         run G1b (table \ref{tab:runs2}).}
\label{fig:conavg_negdudx1000}
\end{figure}

The interactions further from the wall can be investigated by repeating the conditional averaging at ${\bm x}_\times/\delta = (0, 0, 0.5)$. A different relationship between the large and small scale activity emerges, as shown in figure~\ref{fig:conavg_negu1000}.  Instead of the $0$--$\upi/2$ phase difference close to the wall, $|u_{L}^+ - U^+|$ and $|\widetilde{u}_S^+ - \langle\widetilde{u}_S^+\rangle|$ are substantially out of phase, that is the phase difference is approximately $\upi$. This variation is in good agreement with boundary layer results over a Reynolds number range of three decades, namely \cite{Bandyopadhyay1984} and the recent work of \cite{Guala09mod}.  From these other works, it would be expected that the large and small scales have close to zero phase difference very close to the wall; the LES formulation prevents us from confirming this point. %\bjm{The streamwise scale of coherence appears to be shortened} \dc{by how much?} \bjm{in the streamwise direction for this conditional average, perhaps indicating a different structural origin. We note that the recent work of \cite{Monty2009} has indicated the increasing energetic importance of scales of order $3 \delta$ further from the wall than the very large scale energy peak.}

A visual inspection of figure \ref{fig:conavg_negu} gives an estimate that ${\bm u}_L^+$ is of the order $(0.8, 0.2, 0.05)$.
Thus, the shear stress carried directly by these large-scale
structures is estimated from this conditionally averaged picture as $-\langle u_L w_L\rangle^+ \approx 0.04$,
and the large-scale streamwise intensity is estimated as
$\langle u_L^2\rangle^+ \approx 0.64$.  The relative magnitude and phase of the other velocity components associated with the large scale structure can be confirmed by looking at equivalent conditional averages for the other velocity components. Note that it is extremely difficult to obtain this sort of data experimentally, so we are effectively using the LES data in a predictive capacity to complete the description of the trends in the structure of the large-scale motions.

\begin{figure}
\centerline{\includegraphics{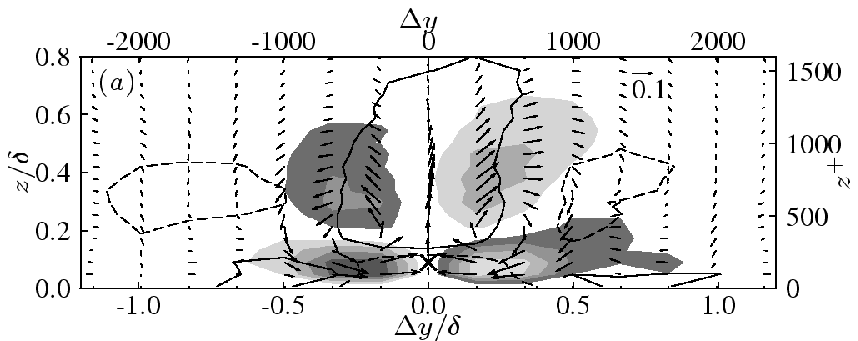}}
\centerline{\includegraphics{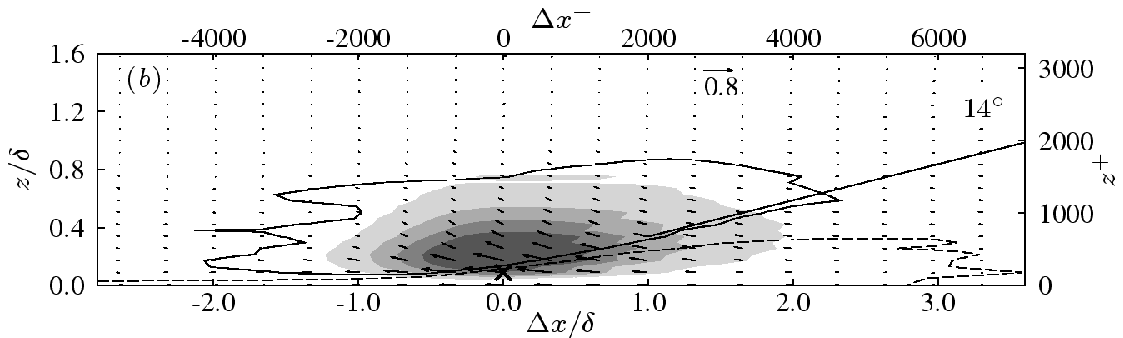}}
\centerline{\includegraphics{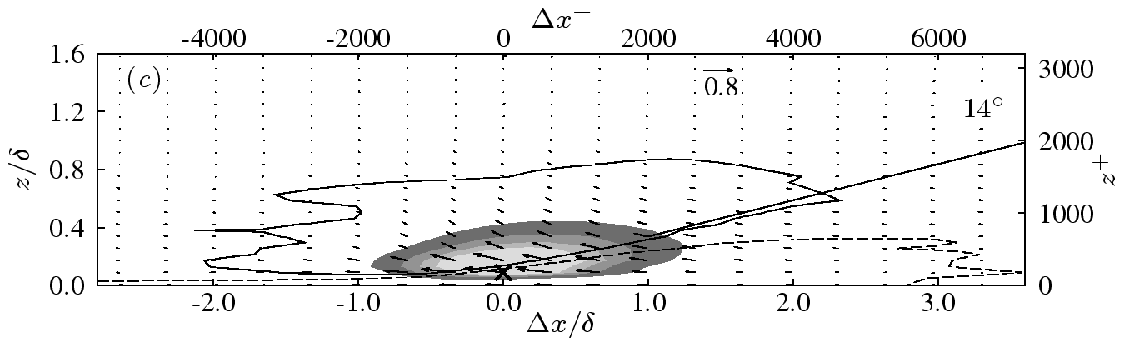}}
\caption{Spanwise and wall-normal velocities, and Reynolds stress associated with the conditionally averaged streamwise velocity in the
%Conditionally-averaged fields in the
(\textit{a}) spanwise--wall-normal and  (\textit{b}, \textit{c}) streamwise--wall-normal planes, with
         $\langle {\bm u}_L^+ | A_1\rangle(\Delta {\bm x})$ and
         $\langle \widetilde{u}_S^+|A_1\rangle(\Delta {\bm x})$,
         where $A_1$ is the low-speed event $u_L({\bm x}_\times) - U < 0$
         with probability $P\{A_1\}=0.53$ at
         ${\bm x}_\times/\delta = (0, 0, 0.090)$:
         (\textit{a}) filled contours of $|\langle\widetilde{v}_L^+|A_1\rangle|
         =0.04,0.08,0.12,0.16$;
         (\textit{b}) filled contours of $|\langle\widetilde{w}_L^+|A_1\rangle|
         =0.03,0.06,0.09,0.12$;
         (\textit{c}) filled contours of $|\langle\widetilde{w}_L^+|A_1\rangle
         (\langle \widetilde{u}_L^+|A_1\rangle-U^+)| = 0.02,0.04,0.06,0.08$; and
         (\textit{a}--\textit{c}) line contours of
          $|\langle \widetilde{u}_S^+|A_1\rangle
           - \langle\widetilde{u}_S^+\rangle| = 0.01$.
         Darker to lighter shades and dashed lines for negative values.
         Vectors represent in-plane velocity components.
         Data from snapshot of $\Rey_\tau = 2\,{\rm k}$ channel flow LES,
         run G1b (table \ref{tab:runs2}).}
\label{fig:conavg_negu_vwwu}
\end{figure}

\begin{figure}
\centerline{\includegraphics{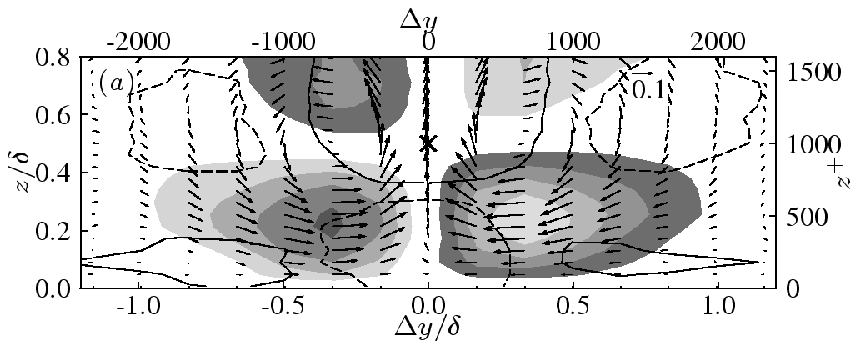}}
\centerline{\includegraphics{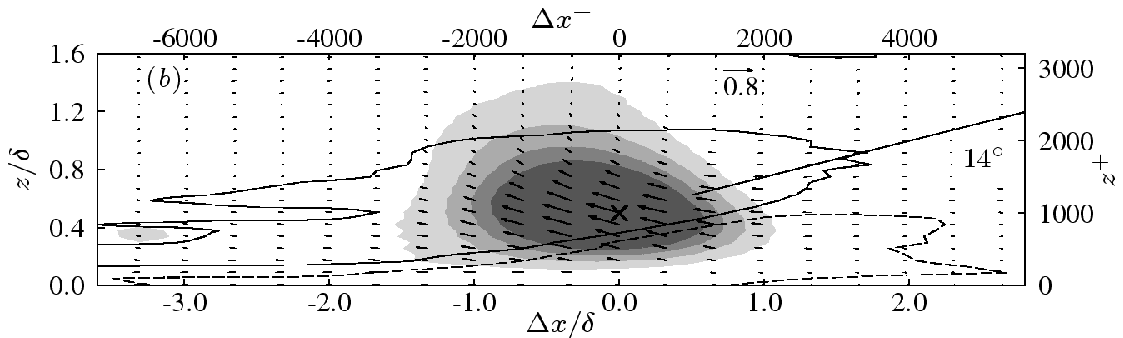}}
\centerline{\includegraphics{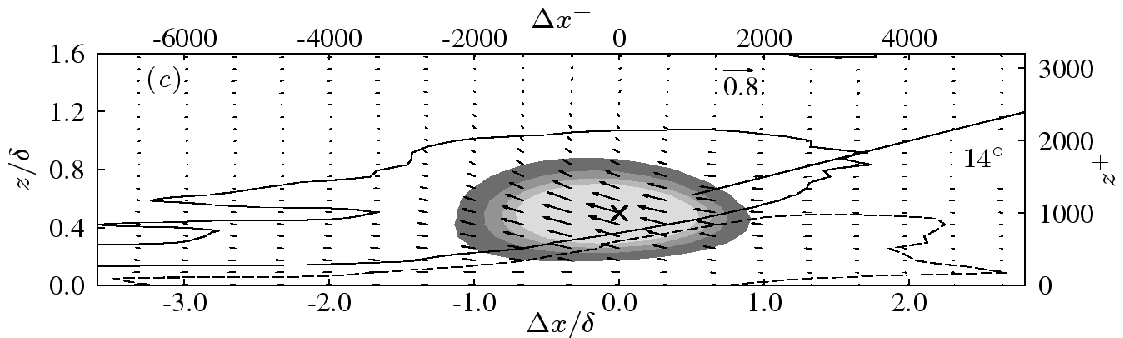}}
\caption{Spanwise and wall-normal velocities, and Reynolds stress associated with the conditionally averaged streamwise velocity in the
%Conditionally-averaged fields in the
(\textit{a}) spanwise--wall-normal and  (\textit{b}, \textit{c}) streamwise--wall-normal planes, with
         $\langle {\bm u}_L^+ | A_2\rangle(\Delta {\bm x})$ and
         $\langle \widetilde{u}_S^+|A_2\rangle(\Delta {\bm x})$,
         where $A_2$ is the low-speed event
         $u_L({\bm x}_\times) - U < 0$
         with probability $P\{A_2\}=0.47$ at
         ${\bm x}_\times/\delta = (0, 0, 0.5)$:
         (\textit{a}) filled contours of $|\langle\widetilde{v}_L^+|A_1\rangle|
         =0.04,0.08,0.12,0.16$;
         (\textit{b}) filled contours of $|\langle\widetilde{w}_L^+|A_1\rangle|
         =0.03,0.06,0.09,0.12$;
         (\textit{c}) filled contours of $|\langle\widetilde{w}_L^+|A_1\rangle
         (\langle \widetilde{u}_L^+|A_1\rangle-U^+)| = 0.02,0.04,0.06,0.08$; and
         (\textit{a}--\textit{c}) line contours of
          $|\langle \widetilde{u}_S^+|A_1\rangle
           - \langle\widetilde{u}_S^+\rangle| = 0.01$.
         Darker to lighter shades and dashed lines for negative values.
         Vectors represent in-plane velocity components.
         Data from snapshot of $\Rey_\tau = 2\,{\rm k}$ channel flow LES,
         run G1b (table \ref{tab:runs2}).}
\label{fig:conavg_negu1000_vwwu}
\end{figure}

\begin{figure}
\centerline{\includegraphics{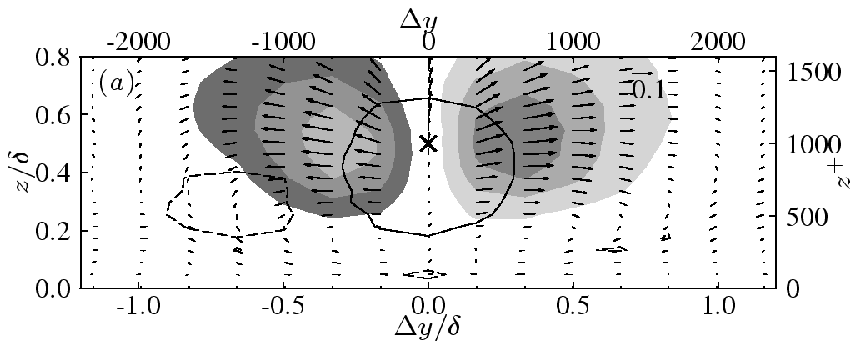}}
\centerline{\includegraphics{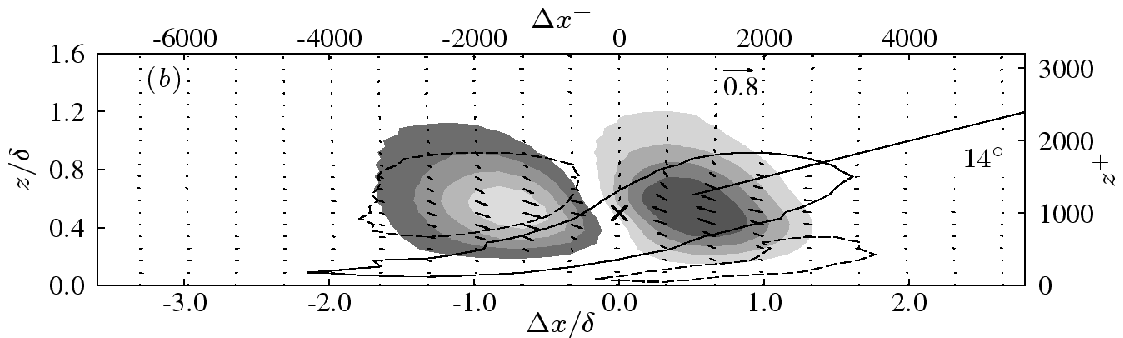}}
\centerline{\includegraphics{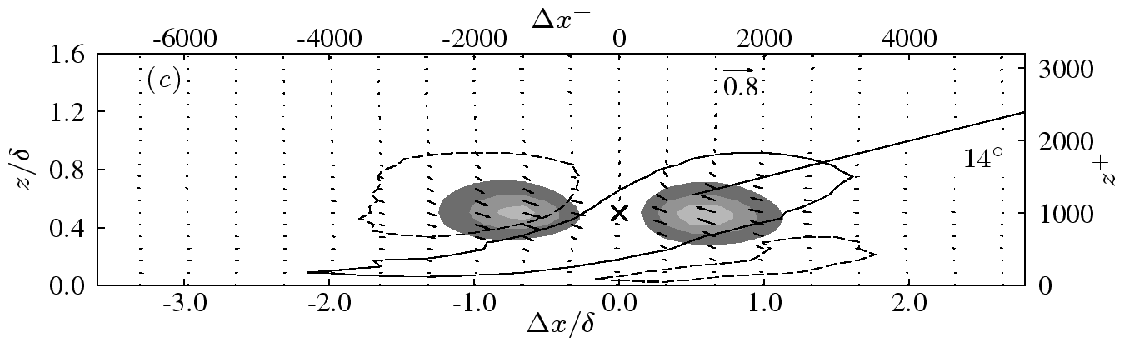}}
\caption{Spanwise and wall-normal velocities, and Reynolds stress associated with the conditionally averaged streamwise velocity in the
%Conditionally-averaged fields in the
         (\textit{a}) spanwise--wall-normal and  (\textit{b}, \textit{c}) streamwise--wall-normal planes, with
         $\langle {\bm u}_L^+ | A_4\rangle(\Delta {\bm x})$ and
         $\langle \widetilde{u}_S^+|A_4\rangle(\Delta {\bm x})$,
         where $A_4$ is the streamwise high-speed-to-low-speed boundary event
         $(\p u_L/\p x)({\bm x}_\times) < 0$
         with probability $P\{A_4\}=0.48$ at
         ${\bm x}_\times/\delta = (0, 0, 0.5)$:
         (\textit{a}) filled contours of $|\langle\widetilde{v}_L^+|A_1\rangle|
         =0.04,0.08,0.12,0.16$;
         (\textit{b}) filled contours of $|\langle\widetilde{w}_L^+|A_1\rangle|
         =0.03,0.06,0.09,0.12$;
         (\textit{c}) filled contours of $|\langle\widetilde{w}_L^+|A_1\rangle
         (\langle \widetilde{u}_L^+|A_1\rangle-U^+)| = 0.02,0.04,0.06,0.08$; and
         (\textit{a}--\textit{c}) line contours of
          $|\langle \widetilde{u}_S^+|A_1\rangle
           - \langle\widetilde{u}_S^+\rangle| = 0.01$.
         Darker to lighter shades and dashed lines for negative values.
         Vectors represent in-plane velocity components.
         Data from snapshot of $\Rey_\tau = 2\,{\rm k}$ channel flow LES,
         run G1b (table \ref{tab:runs2}).}
\label{fig:conavg_negdudx1000_vwwu}
\end{figure}

Figures~\ref{fig:conavg_negu_vwwu}--\ref{fig:conavg_negdudx1000_vwwu}(\textit{a}, \textit{b}) reveal that while the spanwise velocity exhibits a footprint consistent with the implied swirl of figure~\ref{fig:conavg_negu}, such that $v_L = 0$ statistically on the conditioning plane, but with a wall-normal phase variation similar to that exhibited by the streamwise velocity $u_L$. By contrast, the wall-normal velocity velocities in figures~\ref{fig:conavg_negu_vwwu}--\ref{fig:conavg_negdudx1000_vwwu}(\textit{c}) are substantially in phase in the wall-normal direction, independent of the conditioning criterion.

The conditional averaging technique can be taken one step further to demonstrate that, despite a large wall-normal footprint in the streamwise velocity, the large scales will contribute locally to the mean shear stress because of the relative wall-normal phases of the large-scale streamwise and wall-normal velocities.  Figures \ref{fig:conavg_negu_vwwu}--\ref{fig:conavg_negdudx1000_vwwu}(c) show the conditioned shear stress distributions associated with the large scales, that is the product of the conditional averages of $u_L$ and $w_L$.  A conditionally averaged $u_L w_L$ yields a different answer that includes the effects of smaller structures since
$\langle u_L w_L | A_1\rangle \ne \langle u_L | A_1\rangle \langle w_L | A_1\rangle$.
The relative phases of $u_L$ and $w_L$ imply that there will be a contribution to the mean shear stress close to the wall. Note that this result is subject to the success of the conditional averaging in capturing streamwise coherence in $u_L$ and $w_L$, but the trend is believed to be robust.

%\begin{figure}
%\caption{Conditionally-averaged fields in the spanwise--wall-normal and streamwise--wall-normal planes, with
%         $\langle {\bm u}_L^+ | A_1\rangle(\Delta {\bm x})$ and
%         $\langle \widetilde{u}_S^+|A_1\rangle(\Delta {\bm x})$,
%         where $A_1$ is the low-speed event $u_L({\bm x}_\times) - U < 0$
%         with probability $P\{A_1\}=0.53$ at
%         ${\bm x}_\times/\delta = (0, 0, 0.090)$:
%         (\textit{a}, \textit{b}) filled contours of
%         $|u_{L}^+ w_{L}^+| = *******$,
%         line contours of
%         $|\widetilde{u}_S^+ - \langle\widetilde{u}_S^+\rangle| = 0.01$ (darker to lighter shades and dashed
%         lines for negative values, vectors represent in-plane velocity components); and
%         (\textit{c}) streamwise averaged profile of $u_{L}^+ w_{L}^+$.
%         Data from snapshot of $\Rey_\tau = 2\,{\rm k}$ channel flow LES,
%         run G1b (Table \ref{tab:runs2}).}
%\label{fig:conavg_negu_uw}
%\end{figure}

\section{Summary and Conclusions}
\label{sec:conclusions}

%We have reported on the phase relationship between large- and small-scale events in LES of channel flow in long $L_x \approx 96\delta$ channels. %While the approximations inherent in LES have been acknowledged at the outset, 
We have designed a series of LES runs that are well-suited to the investigation of large-scale structures in a long channel.
%These structures appear to be robust to details at the smaller scales.
The observations from this study lend themselves to an interpretation of the
ubiquitous influence of large-scale structures in turbulent channel flow,
consistent with the boundary layer experiments of
\cite{Bandyopadhyay1984, Mathis2009, Guala09mod}.
%The observations from this study are qualitatively consistent with the boundary layer experiments of \cite{Bandyopadhyay1984, Mathis2009, Guala09mod} and lend themselves to an interpretation of the ubiquitous influence of large-scale structures in turbulent channel flow.

Using simultaneous time and spatial data to construct the spatio-temporal
spectrum, we compute the dispersion relation, and show
that departure from Taylor's frozen-turbulence hypothesis
is noticeable near the wall, $z/\delta<0.08$, and for
large scales, $>6\delta$, in channel flow at $\Rey_\tau = 2\,{\rm k}$. This is consistent with the footprint of the very large scales reaching down to the wall.
The opposite effect---that the large scales convect slower than the
local mean---is also observed for these large scales away
from the wall ($z/\delta=0.5$).
This, too, is consistent with the conditional averages that show
that these large-scale structures reach far away from the wall.
%\bjm{Slight but visible effect in the core of the pipe.}
%Top of the large-scale structure showing convection velocity slower than the mean flow -- conditional averages certainly show that the dominant structure reaches far from the wall

Flow fields constructed from conditional averages confirm the extent of the influence of scales with $\lambda_x \approx 6\delta$ and reveal the wall-normal dependence of the spatial relationship between the high-speed large-scale region 
and the underlying small-scale intensity. The LES results appear to underline that the apparent amplitude modulation effect is better described in terms of the spatial phase between the large and the small scales; indeed our correlation coefficients, $R_\rho$ and $R_\tau$, are formulated in terms of this phase relationship. In this context, the zero in the correlation between large and small scale activity described in \S\,\ref{sect:R} can be interpreted as the location where, on average, the corresponding signals are $\upi/2$ out of phase. A picture emerges in which the small scales are nominally in phase with the large scales near the wall and $\upi$ out of phase further from the wall. In the intervening region, where the phase difference is approximately $\upi/2$, the small scales track the sign of $\p u_L/\p x$, per the small-scale production/backscatter
term $-2\widetilde{u}_S^2\p u_L/\p x$.  We can interpret the small scale activity, $\widetilde{u}_S$, in terms of local structure, noting that the inferred locus of the maximum small scale energy corresponds to the well-known structure inclination angle of approximately $10$--$20^\circ$, at least at this Reynolds number.

We conclude by observing that the very large scales appear to dictate some of the turbulence behavior close to the wall, with interesting implications for the ``top-down versus bottom-up'' debate concerning the dynamical significance of different regions of the flow.  It is perhaps more correct to say simply that the two regions can now be determined to be inextricably linked.% \dc{True, but one can still attribute the larger part of the Reynolds shear and normal stress budget solely to small scales.}

%VLSM peak cf R cross-over?

%skewness comments on small vs large scales.

\begin{acknowledgments}
The authors acknowledge the support of the NSF under grants CBET-0651754 (D.C.) and CAREER-0747672 (B.J.M.).
It is our pleasure to acknowledge Professor D.\,I.\,Pullin's
generous support for this investigation, and useful discussions with Drs. Ati Sharma and Michele Guala.

\end{acknowledgments}

\bibliographystyle{jfm}
% Note the spaces between the initials
\bibliography{wallstruct09v1.3}

\end{document}